\documentclass[a4paper,11pt]{article}
\pdfoutput=1 

\usepackage{jheppub} 

\usepackage[T1]{fontenc} 
\usepackage{slashed,bbm,here}

\title{\boldmath Anomalous dimensions of higher spin currents in large $N$ CFTs}

\author[a,b]{Yasuaki Hikida}
\author[b]{Taiki Wada}

\affiliation[a]{Center for Gravitational Physics, Yukawa Institute for Theoretical Physics, Kyoto University,\\ Kyoto 606-8502, Japan}

\affiliation[b]{Department of Physical Sciences, College of Science and Engineering, Ritsumeikan University,\\Shiga 525-8577, Japan}

\emailAdd{yhikida@yukawa.kyoto-u.ac.jp}
\emailAdd{rp0017xp@ed.ritsumei.ac.jp}

\abstract{We examine anomalous dimensions of higher spin currents in the critical O$(N)$ scalar model and the Gross-Neveu model in arbitrary $d$ dimensions. These two models are proposed to be dual to the type A and type B Vasiliev theories, respectively. We reproduce the known results on the anomalous dimensions to the leading order in $1/N$ by using conformal perturbation theory. This work can be regarded as an extension of previous work on the critical O$(N)$ scalars in 3 dimensions, where it was shown that the bulk computation for the masses of higher spin fields on AdS$_4$ can be mapped to the boundary one in conformal perturbation theory. The anomalous dimensions of the both theories agree with each other up to an overall factor depending only on $d$, and we discuss the coincidence for $d=3$ by utilizing $\mathcal{N}=2$ supersymmetry.

}

\arxivnumber{1610.05878}
\preprint{YITP-16-117}

\begin{document}
	\maketitle
	\flushbottom

\section{Introduction}

Higher spin gauge theory is expected to describe superstring theory by breaking its symmetry \cite{Gross:1988ue}, and it is important to understand the mechanism of symmetry breaking in higher spin gauge theory.
In AdS space higher spin gauge theory with consistent interaction has been constructed in \cite{Vasiliev:1995dn,Vasiliev:1999ba,Vasiliev:2003ev}, and the theory can be examined from dual CFT by making use of AdS/CFT correspondence.
In \cite{Hikida:2016wqj},  the anomalous dimensions of higher spin currents in the 3d critical O$(N)$ scalar model were reproduced to the leading order in $1/N$ using conformal perturbation theory. 
The 3d O$(N)$ scalar model was proposed to be dual to so-called type A Vasiliev theory on AdS$_4$ \cite{Klebanov:2002ja},  and the mass $M_s$ of spin $s$ field on AdS$_4$ can be read off from the scaling dimension $\Delta_s$ of dual current as
\begin{align}
M^2_{s} = \Delta_{s} (\Delta_{s} - d) - (d+ s - 2) ( s -2) 
\label{dictionary}
\end{align}
with $d=3$.
The computation in conformal perturbation theory can explain some of bulk interpretations on the symmetry breaking.

The Vasiliev theory dual to the 3d O$(N)$ scalars has been examined since it is the most fundamental example, but more extended versions should be used in order to see the relation to superstring theory. For examples, it was conjectured in \cite{Chang:2012kt} that a 4d extended Vasiliev theory is related to superstring theory on AdS$_4 \times$CP$^3$ via ABJ(M) theory \cite{Aharony:2008ug,Aharony:2008gk}. Moreover, it was also discussed that 3d extended Vasiliev theories are related to superstring theories on AdS$_3$ via $\mathcal{N}=4$ supersymmetric 2d models
\cite{Gaberdiel:2013vva,Gaberdiel:2014cha}  and via $\mathcal{N}=3$ supersymmetric 2d models  \cite{Creutzig:2014ula,Hikida:2015nfa}.
For the relation to AdS$_3$ strings, the symmetry breaking has been examined in \cite{Hikida:2015nfa,Gaberdiel:2015uca,Creutzig:2015hta}. See also \cite{Gwak:2015jdo,Gwak:2016sma} for related works.

In this paper, we first extend the analysis in \cite{Hikida:2016wqj} to the critical O$(N)$ scalar model in arbitrary $d$ dimensions,
which is supposed to be dual to the type A Vasiliev theory on AdS$_{d+1}$ \cite{Klebanov:2002ja}.
We then apply the methods to the Gross-Neveu model  \cite{Gross:1974jv,Moshe:2003xn} in $d$ dimensions, where the model is given by a theory of fermions with four fermi interactions.
The model is proposed to be dual to the type B Vasiliev theory in \cite{Leigh:2003gk,Sezgin:2003pt}, but the equations of motion are known only for $d=3$.
In order to apply our analysis to the conjectured triality of \cite{Chang:2012kt}, we need to introduce supersymmetry and the coupling to Chern-Simons gauge fields in the CFTs as discussed in \cite{Aharony:2011jz,Giombi:2011kc,Maldacena:2012sf}.
One of the main motivations to examine the Gross-Neveu model is for the preparation of the application.

The system of free O$(N)$ real scalars $\phi^i$ $(i=1,2,\cdots,N)$ has a relevant deformation of the double-trace type as 
\begin{align}
 \Delta S = \frac{f}{2} \int d^d x \mathcal{O}_\nu (x) \mathcal{O}_\nu (x) \, , 
 \label{def}
\end{align}
where $\mathcal{O}_\nu (x)$ is a scalar operator with the scaling dimension $\nu$. In the current case, we set $\mathcal{O}_{d-2} = \phi^i \phi_i$.
The critical O$(N)$ scalar model with $d < 4$ can be realized as a CFT at an IR fixed point of the RG flow induced by the deformation, and we study the critical model by treating the deformation perturbatively.
The Gross-Neveu model with $2 < d < 4$ can be also realized in a similar manner. 
The system of free U$(\tilde N)$ free Dirac fermions $\psi^i$ $(i=1,2,\cdots ,\tilde N)$ has an irrelevant deformation of the form \eqref{def} with $\mathcal{O}_{d-1} = \bar \psi^i \psi_i$.
The Gross-Neveu model can be realized as a CFT at a UV fixed point, and we study the model in conformal perturbation theory.

The Vasiliev theory includes scalar field along with higher spin gauge fields, and boundary condition for the scalar field should be specified. 
The free O$(N)$ scalar theory is dual to the type A Vasiliev theory with Neumann boundary condition for the bulk scalar. Since the double-trace type deformation is known to be dual to the change of boundary condition \cite{Witten:2001ua}, the IR fixed point is dual to the Vasiliev theory with Dirichlet boundary condition assigned to the scalar field. 
Similarly, the free fermion theory and the Gross-Neveu model are dual to the type B Vasiliev theories with Dirichlet and Neumann boundary conditions, respectively. 
The higher spin symmetry is broken in the critical models to the $1/N$ order due to the deformation.
This implies that the bulk higher spin gauge symmetry should be broken in a one-loop effect and due to the change of boundary condition for a scalar field \cite{Girardello:2002pp}.
In \cite{Hikida:2016wqj}, additional supports were provided by mapping the computation of anomalous dimensions in the 3d O$(N)$ scalar model in conformal perturbation theory to the evaluation of bulk Witten diagram corresponding to the one-loop effect generating the masses of dual higher spin fields on AdS$_4$.%
\footnote{The essential part of map is rewriting the one-loop Witten diagram as a product of tree ones as  was already proposed in \cite{Hartman:2006dy,Giombi:2011ya}.}
We do not repeat in this paper but it is a straightforward task to generalize the map to our cases.
Thus the same conclusion can be applied to the Vasiliev theories on AdS$_{d+1}$ dual to the critical O($N$) scalars and the Gross-Neveu model in $d$ dimensions.

This paper is organized as follows;
In the next section, we explain the general methods as developed in \cite{Hikida:2016wqj} but with a difference  on the regularization scheme. In this paper, we adopt the one developed in  \cite{Vasiliev:1975mq,Vasiliev:1981yc,Vasiliev:1981dg,Derkachov:1997ch,Diab:2016spb},
which seem to be a common way to regularize Feynman integrals in $1/N$-expansions and in general dimensions.
In section \ref{scalar}, we apply the methods to the O$(N)$ scalar model in $d$ dimensions, and we reproduce the known results on anomalous dimensions of higher spin currents \eqref{anomalouss} obtained in \cite{Lang:1992zw} (see also \cite{Skvortsov:2015pea,Giombi:2016hkj} for recent developments).
In section \ref{GN}, we examine on the Gross-Neveu model in $d$ dimensions.
With a slight modification of the method, we derive the anomalous dimensions of higher spin currents, which agree with the results in \cite{Muta:1976js}. They are the same as those of scalar case up to a factor depending only on $d$, see \eqref{anomalousf}. 
In section \ref{General}, we study simple generalizations of our results.
In particular, we examine the coincidence of anomalous dimensions in theories of bosons and fermions for $d=3$ by making use of $\mathcal{N}=2$ supersymmetry.
In section \ref{conclusion},we conclude this paper and discuss open problems.
In appendices \ref{alt} and \ref{technical} we collect technical materials.

\section{Methods}
\label{Methods}

Let us start from free theory with scalars or fermions in $d$ dimensions.
The theory includes conserved currents $J_{\mu_1 \cdots \mu_s} (x)$ with symmetric and traceless indices. The free fermion theory in $d > 3$ dimensions has mixed symmetry currents as well, but we do not consider them in this paper.
It is convenient to express the currents as $ J^\epsilon_s (x) =J_{\mu_1 \cdots \mu_s} (x) \epsilon^{\mu_1} \cdots \epsilon ^ {\mu_s}$ with polarization vector $\epsilon$ $(\epsilon \cdot \epsilon = 0)$.  The conserved current has scaling dimension $d - 2 + s$, and  the current-current two point function is fixed by symmetry as
\begin{align}
 \langle J_s^\epsilon (x_1) J_s ^\epsilon (x_2) \rangle = N_s \left( \frac{\hat x_{12}}{|x_{12}|} \right)^{2s} \frac{1}{(x_{12}^2)^{d - 2 +  s}} = N_s \frac{(\hat x_{12})^{2s}}{(x_{12}^2)^{d-2 + 2 s}}
\label{freeJs2pt}
\end{align}
up to an overall factor $N_s$. Here we have expressed $\hat x_{12} = \epsilon \cdot (x_1 - x_2)= \epsilon \cdot x_{12}$.

We deform free theories by double-trace type deformation as in \eqref{def}.
The scalar two point function is given by
\begin{align}
\langle \mathcal{O}_\nu (x_1) \mathcal{O}_\nu (x_2) \rangle =  \frac{C_{\mathcal{O}_\nu}}{(x_{12}^2)^{\nu}}
\label{scalar2pt}
\end{align}
with a normalization factor $C_{\mathcal{O}_\nu}$.
We compute correlation functions in the presence of  deformation using the conformal perturbation theory as
\begin{align}
\left \langle \prod_{i=1}^n A_i (x_i) \right \rangle _f 
= \frac{\langle \prod_{i=1}^n A_i (x_i) e^{- \Delta S} \rangle_0 }{\langle e^{- \Delta S} \rangle_0} 
\end{align}
with some operators $A_i$. Here the right hand side is computed in the free theory.

As an important example, we examine how the scalar two point function in \eqref{scalar2pt} changes under the deformation \eqref{def}.
It is convenient to work in the momentum representation as
\begin{align}
\mathcal{O}_\nu(p) = \frac{1}{(2 \pi)^{d/2}} \int d^d x \mathcal{O} _\nu (x) e^{- i p \cdot x} \, ,
\end{align}
which leads to
\begin{align}
F_\nu(p) \equiv \langle \mathcal{O}_\nu (p) \mathcal{O}_\nu (-p) \rangle_0 = \frac{C_{\mathcal{O}_\nu} \pi^{d/2 } 2^{d- 2 \nu} a(\nu) }{(p^2)^{d/2 - \nu}} \, , \quad 
\Delta S = \frac{f}{2}\int d^d p  \mathcal{O}_\nu(p) \mathcal{O}_\nu(-p) \, .
\end{align}
Here we have used the formula for the Fourier transformation as
\begin{align}
\frac{1}{(x^2)^{ \alpha}} = \frac{a(\alpha)}{\pi^{d/2} 2^{2 \alpha}} \int d ^d p \frac{e^{i p \cdot x}}{(p^2)^{d/2 - \alpha}} \, , \quad a (\alpha) = \frac{\Gamma (d/2 - \alpha)}{\Gamma(\alpha)} \, .
\label{Fourier}
\end{align}
After the deformation, the two point function becomes
\begin{align}
\langle \mathcal{O}_\nu (p) \mathcal{O}_\nu (-p) \rangle _{f}
 = F_\nu(p) - f F_\nu(p)^2 + f^2 F_\nu (p)^3 - \cdots = \frac{F_\nu (p)}{1 + f F_\nu(p)} \, .
\label{summedup}
\end{align}
We will be interested in the regime with $f \sim \infty$, where we find
\begin{align}
\langle \mathcal{O}_\nu (p) \mathcal{O}_\nu (-p) \rangle _{f} \sim \frac{1}{f}- \frac{1}{f^2} F_\nu(p)^{-1} \, ,
\end{align}
or in the coordinate representation
\begin{align}
\langle \mathcal{O}_\nu (x_1) \mathcal{O}_\nu (x_2) \rangle _{f} \sim \frac{1}{f} \delta^{(d)} (x_{12})  + \frac{1}{f^2} G_\nu (x_{12}) 
\end{align}
with 
\begin{align}
G_\nu (x_{12}) =   -  
\frac{a(\nu - d/2)}{C_{\mathcal{O}_\nu} \pi^d a (\nu)} \frac{1}{(x_{12}^2)^{d - \nu}} \, .
\label{Gx12}
\end{align}
Therefore, we can see that the deformation changes the dimension of  operator $\mathcal{O}_\nu (x)$ from $\nu$ to $d - \nu$.

With the deformation, the higher spin current would have anomalous dimension $\gamma_s$,
which is defined by
\begin{align}
\gamma_s \equiv \Delta_s - (d -2 + s) 
\label{gammas}
\end{align}
with $\Delta_s$ as the scaling dimension of spin $s$ current.
We would like to obtain $\gamma_s$ at $f \to \infty$, and
for the purpose, we compute the two point function of higher spin current at $f \to \infty$ using the conformal perturbation theory.
The form of the two point function is fixed  as
\begin{align}
 \langle J_s^\epsilon (x_1) J_s ^\epsilon (x_2) \rangle _{f \to \infty}& = N_s (1 + \delta_s) \frac{(\hat x_{12})^{2s}}{(x_{12}^2)^{d - 2 + 2 s + \gamma_s}} \nonumber \\
 &=  N_s  \frac{(\hat x_{12})^{2s}}{(x_{12}^2)^{d - 2 + 2 s }} (1 + \delta_s - \gamma_s \log (x_{12}^2)) + \cdots \, ,
\end{align}
where the shift of normalization $\delta_s$ is of order $1/N$. Therefore, we can read off the anomalous dimensions from the term proportional to $\log (x_{12}^2) $.

The conformal perturbation theory says that the two point function should be computed as
\begin{align}
&\langle  J_s^\epsilon (x_1) J_s^\epsilon (x_2) \rangle_{f } = 
\langle  J_s^\epsilon (x_1) J_s^\epsilon (x_2) \rangle_{0} 
- \frac{f}{2} \int d^d x_3 \langle  J_s^\epsilon (x_1) J_s^\epsilon (x_2) \mathcal{O}_\nu (x_3) \mathcal{O}_\nu (x_3) \rangle_{0}  \\
&+ \frac{f^2}{8}  \int d^d x_3 d^d x_4 \langle  J_s^\epsilon (x_1) J_s^\epsilon (x_2) \mathcal{O}_\nu (x_3) \mathcal{O}_\nu (x_3)  \mathcal{O}_\nu (x_4) \mathcal{O}_\nu (x_4) \rangle_{0} + \cdots  \nonumber \, .
\end{align}
We would like to compute the anomalous dimensions to the leading order in $1/N$.
With the large $N$ factorization, the correlators  can be written as products of three or four point functions involving $J^\epsilon_s$ and two point functions of $\mathcal{O}_\nu$. The product of scalar two point functions can be summed up as in \eqref{summedup}.
As a result we have at the leading order in $1/N$
\begin{align}
&\langle  J_s^\epsilon (x_1) J_s^\epsilon (x_2) \rangle_{f \to \infty} 
 =\langle J_s^\epsilon (x_1) J_s^\epsilon (x_2) \rangle_{0}  + I_1  + I_2 + \cdots 
 \end{align}
with
 \begin{align}
 \label{integrals}
 &I_1 =  \frac12 \int d^d x_3 d^d x_4 G_\nu(x_{34}) 
   \langle J_s^\epsilon (x_1) J_s^\epsilon (x_2) \mathcal{O}_\nu(x_3) \mathcal{O}_\nu(x_4) \rangle_{0} \, , \\
 & I_2 = \frac12 \int d^d x_3 d^d x_4  d^d x_5 d^d x_6 G_\nu(x_{35}) G_\nu(x_{46}) 
      \langle J_s^\epsilon(x_1)  \mathcal{O}_\nu(x_3) \mathcal{O}_\nu(x_4) \rangle_{0}
       \langle J_s^\epsilon (x_2)  \mathcal{O}_\nu(x_5) \mathcal{O}_\nu(x_6) \rangle_{0} \, . \nonumber
\end{align}
Here $G_\nu (x_{12})$ is defined in \eqref{Gx12} and dots denote sub-leading terms in $1/N$.
For more detailed derivation, see \cite{Hikida:2016wqj}.
The same expression can be obtained by introducing a Hubbard-Stratonovich auxiliary field $\sigma$ as in \cite{Diab:2016spb}. 
In the latter formulation, $G_\nu (x_{12})$ can be regarded as the effective propagator of $\sigma$, and the interaction becomes $\Delta S = \int d^d x \sigma \mathcal{O}_\nu $. 

Since the integrals diverge, we need to adopt a way to regularize them.
In \cite{Hikida:2016wqj}, we used a dimensional regularization, which reproduces the known result.
In order to work with arbitrary real dimension $d$, we adopt a different regularization by shifting the dimension of the auxiliary field $\sigma$, which seems to be a more common way \cite{Vasiliev:1975mq,Vasiliev:1981yc,Vasiliev:1981dg,Derkachov:1997ch,Diab:2016spb}.
In the regularization scheme, we change the propagator \eqref{Gx12} as
\begin{align}
G_\nu(x_{12} ) =  -  
\frac{a(\nu - d/2)}{C_{\mathcal{O}_\nu} \pi^d a (\nu)} \frac{1}{(x_{12}^2)^{d - \nu - \Delta}} \, .
\label{Greg}
\end{align}
This shift makes the coupling for the interaction dimensionful, thus we may write $\Delta S = \mu^\Delta \int d^d x \sigma \mathcal{O}$ with a regularization scale $\mu$.
Denoting the residues of $I_1$ and $I_2$ in \eqref{integrals} at $\Delta = 0$ as $I_1^\text{res}$ and $I_2^\text{res}$, the anomalous dimensions can be read off from the term proportional to $\log (x_{12}^2)$ as (see \cite{Diab:2016spb})
\begin{align}
I_1 + I_2 &= (x_{12}^2 \mu ^2 )^\Delta  \left(\frac{1}{\Delta} I_1^\text{res} +  I_1^\text{fin} + \mathcal{O} (\Delta) \right) +  (x_{12} ^2 \mu ^2 )^{2\Delta}  \left(\frac{1}{\Delta} I_2^\text{res} +  I_2^\text{fin} + \mathcal{O} (\Delta) \right)  \\
&= \frac{1}{\Delta} (I_1^\text{res} +  I_2^\text{res}) + \log (x_{12}^2 \mu^2 ) 
\left(I_1^\text{res} +  2 I_2^\text{res} \right) + I_1^\text{fin} +  I_2^\text{fin} + \mathcal{O} (\Delta) \, . \nonumber
\end{align}
In the following sections, we compute $I_1^\text{res}$ and $I_2^\text{res}$ both for the O$(N)$ scalar model and the Gross-Neveu model in $d$ dimensions.

In order to perform  explicit computations, it is convenient to utilize the integral formulas in \cite{Kazakov:1983ns}. 
For the chain integral, we use
\begin{align}
\int d^d x_3 \frac{1}{(x_{13}^2)^{\alpha_1} (x_{23}^2)^{\alpha_2} } 
 = v(\alpha_1 ,\alpha_2 , \alpha_3) \frac{1}{(x_{12}^2)^{\alpha_1 + \alpha_2 - d/2} }\, ,
 \label{chain}
\end{align}
where
\begin{align}
v(\alpha_1 , \alpha_2 , \alpha_3) = \pi^{d/2} \prod_{i=1}^3 a (\alpha_i) \, . 
\end{align}
Here $\alpha_3 = d - \alpha_1 - \alpha_2$ and $a(\alpha)$ is defined in \eqref{Fourier}.
Graphically, it can be represented as
\begin{figure}[H]
 \centering
 \includegraphics[keepaspectratio, scale=0.5]
      {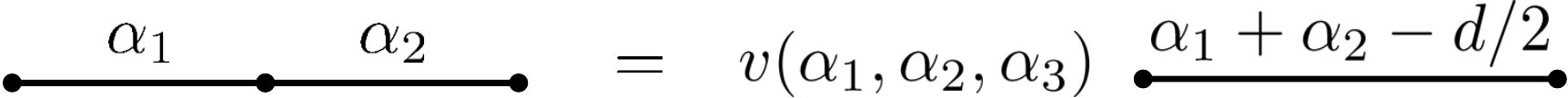}
\end{figure}
\noindent
For the vertex integral with $\alpha_1 + \alpha_2 + \alpha_3 = d$, we can use
\begin{align}
\int d^d x_4 \frac{1}{(x_{14}^2)^{\alpha_1}(x_{24}^2)^{\alpha_2}(x_{34}^2)^{\alpha_3}}
 = v(\alpha_1 , \alpha_2 ,\alpha_3) \frac{1}{(x_{23}^2)^{d/2 - \alpha_1}(x_{31}^2)^{d/2 - \alpha_2}(x_{12}^2)^{d/2 - \alpha_3}} \, ,
 \label{unique}
\end{align}
which is the main element of  integration.
This integral can be also visualized graphically as
\begin{figure}[H]
 \centering
 \includegraphics[keepaspectratio, scale=0.5]
      {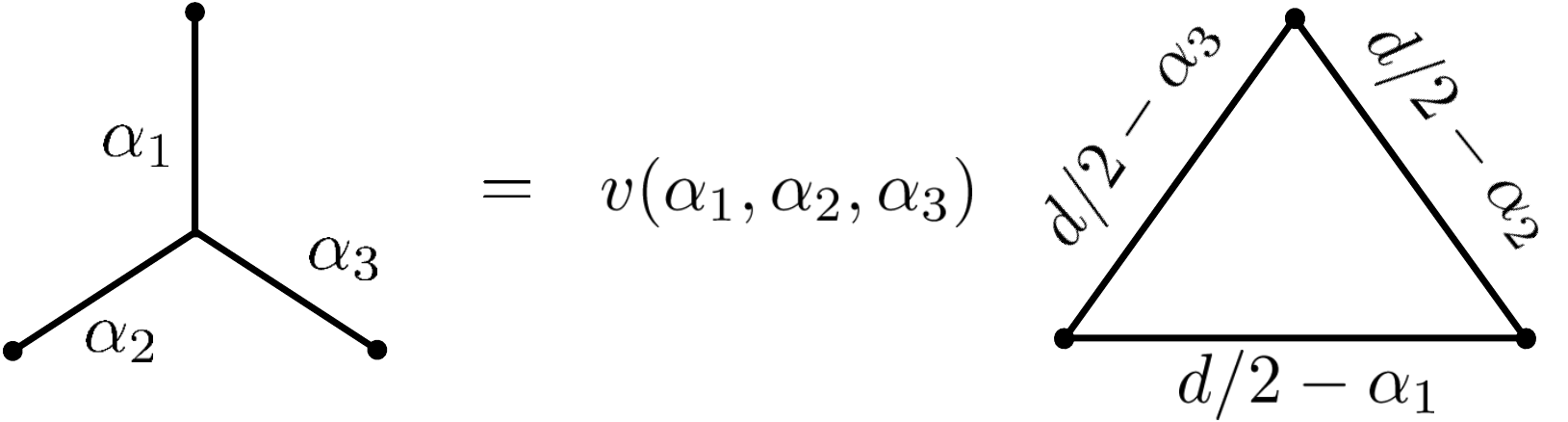}
\end{figure}
\noindent
For the vertex integral with  $\alpha_1 + \alpha_2 + \alpha_3 \neq d$, we may choose to work in the momentum representation by applying the Fourier transform \eqref{Fourier} not in the coordinate representation.

\section{O$(N)$ scalar model}
\label{scalar}

In the previous section, we have extended the methods developed in \cite{Hikida:2016wqj} in order to apply for more generic examples. In particular, we modify them by incorporating the regularization with the shift of exponent as in \eqref{Greg}. 
In this section, we apply the extended methods to examine  the critical O$(N)$ scalar model in arbitrary  dimensions, and we reproduce the anomalous dimensions of higher spin currents, which was obtained in \cite{Lang:1992zw}.

\subsection{Anomalous dimensions of higher spin currents}

We consider the system of $N$ free real scalars  $\phi^i$ $(i=1,2,\cdots , N)$ in $d$ dimensions.
The system has conserved currents $ J^\epsilon_s  (x)$ with even spin $s$.
We use the expression of currents obtained in \cite{Anselmi:1999bb} as%
\footnote{The expression is the same as the one with Gegenbauer polynomial up to an overall factor, see, e.g., \cite{Skvortsov:2015pea,Giombi:2016hkj}.}
\begin{align}
\label{currentdef}
 J_s ^\epsilon = \sum_{k=0}^s a_k \hat \partial^{s-k} \phi^i \hat \partial^k \phi_i \, , \quad
  a_k 
 =  \frac{(-1)^k}{2} \frac{\binom{s}{k} \binom{s+d-4}{k+d/2 - 2}}{\binom{s+d-4}{d/2 - 2}} \end{align}
with $\hat \partial = \epsilon \cdot \partial $.
Correlation functions involving the scalar fields
 can be computed by applying the Wick contraction
\begin{align}
\langle \phi^i (x_1) \phi^j (x_2) \rangle = C_\phi 
\frac{\delta^{ij}}{(x_{12}^2)^{\delta}}\, , \quad C_\phi = \frac{\Gamma (\delta)}{4 \pi^{d/2}}  \, . \label{Wicks}
\end{align}
Here we set $\delta = d/2 - 1$.
Using the Wick contraction, the current-current two point function is obtained as
\begin{align}
 \langle J_s^\epsilon (x_1) J_s ^\epsilon (x_2) \rangle = 2 N C_\phi^2
\sum_{k,l=0}^s a_k a_l 
\left[ \hat \partial^k_1 \hat \partial^{s-l}_2 \frac{1}{(x_{12}^2)^\delta} \right] 
\left[ \hat \partial^l_2 \hat \partial^{s-k}_1 \frac{1}{(x_{21}^2)^\delta} \right]  =
C_s \frac{(\hat x_{12})^{2s}}{(x_{12}^2)^{2 \delta + 2 s}} \, ,
\label{current2pt}
\end{align}
where the coefficient can be computed as \cite{Diaz:2006nm}
\begin{align} 
 C_s  = 
  2^{2s+1}N C_\phi^2 \sum_{k,l=0}^s a_k a_l (\delta)_{k+l} (\delta)_{2s - k - l}
   = C_\phi^2 2^{2s-1}  N s! (2 \delta - 1 + s)_s
\label{current2ptCs}
\end{align}
for $s > 0$. Here we have used the Pochhammer symbol as $(a)_n = \Gamma(a+n)/\Gamma(a)$.
We shall denote the two point function as 
\begin{align}
D_0^s \equiv  \langle J^\epsilon_s (x_1)  J^\epsilon_s (x_2) \rangle   
\label{D0s}
\end{align}
for later use.

We would like to read off anomalous dimensions after the double-trace type deformation in \eqref{def}. In the present case, we deform the theory by using the scalar operator
\begin{align}
\mathcal{O}_{2\delta}(x) \equiv \mathcal{O}(x) = \phi^i \phi_i \, , \quad \langle \mathcal{O}(x_1) \mathcal{O}(x_2) \rangle
 = C_\mathcal{O} \frac{1}{(x_{12}^2)^{2\delta}} \, , \quad 
 C_\mathcal{O} = 2N C_\phi^2  \, .
\end{align}
The operator is dual to the bulk scalar field, and the deformation changes its boundary condition from the Neumann one at $f =0$ to the Dirichlet one at $f = \infty$. 
With this deformation, the propagator \eqref{Greg} becomes
\begin{align}
G_{2 \delta}(x_{12}) \equiv G(x_{12}) = \frac{C_\sigma}{(x_{12}^2)^{2 - \Delta}} \, , \quad
C_\sigma = \frac{(d/2 - 2) \sin (\pi d/2 ) \Gamma (d - 2) }{C_\mathcal{O} \pi^{d+1}} \, .
\label{sigmapro}
\end{align}
With the leading anomalous dimension for $\phi^i$, it can be rewritten as
\begin{align}
C_\sigma = d (d - 2) \gamma_\phi \, , \quad \gamma_\phi = \frac{2 \sin (\pi d /2) \Gamma (d-2) }{N \pi \Gamma (d/2 - 2) \Gamma (d/2 + 1)} \, .
\label{Csigma}
\end{align}
In the rest of this section, we evaluate the residues of integrals $I_1$ and $I_2$ in \eqref{integrals} at $\Delta = 0$.
The results are summarized as
\begin{align}
&I_1^\text{res} = - 2 \gamma_\phi \left(1  -   \frac{d (d-2) }{4 (s + d/2 - 2) (s + d/2 - 1)} \right) D_0^s 
 = - 2 \gamma_\phi \frac{(s-1)(s + d - 2) }{ (s + d/2 - 2) (s + d/2 - 1)} D_0^s  \, , \nonumber \\
&I_2^\text{res} = \gamma_\phi \frac{ \Gamma(d+1) \Gamma( s + 1) }{ 2 (d-1) \Gamma (d + s - 3) ( s + d/2 - 2) (s + d/2 - 1)} D_0^s  \, .
\label{scalarresult}
\end{align}
For $d=3$, they agree with \cite{Hikida:2016wqj}.
For generic $d$ with $s=2$, the same integrals (and  $I_1$ with $s=1$)  have been computed in \cite{Diab:2016spb}  including finite part, and our results reproduce the $1/\Delta$-pole terms. 
The expressions for $s \geq 3$ are new.
The anomalous dimension $\gamma_s$ is then read off as
 \begin{align}
  \label{anomalouss}
 \gamma_s = 2 \gamma_\phi \frac{1}{ (s + d/2 - 2) (s + d/2 - 1)} \left[ (s-1)(s + d - 2) - \frac{ \Gamma(d+1) \Gamma( s + 1) }{ 2 (d-1) \Gamma (d + s - 3) }  \right] \, ,
 \end{align}
which agrees with the known result in \cite{Lang:1992zw}.

\subsection{Integral $I_1$}

We first compute the integral $I_1$. Applying the Wick contraction \eqref{Wicks}, we can see that there are two types of contributions to the four point function as
\begin{align}
 \langle J_s^\epsilon (x_1) J_s^\epsilon (x_2) \mathcal{O}(x_3) \mathcal{O}(x_4) \rangle_{0} = K_1 (x_i) + K_2 (x_i) \, , \label{fourpt}
\end{align}
where
\begin{align}
&K_1 (x_i) = 16N C_\phi^4 \sum_{k,l=0}^s a_k a_l \left[ \hat \partial^k_1 \hat \partial^{s-l}_2 \frac{1}{(x_{12}^2)^\delta} \right] \left[ \hat \partial^l_2 \frac{1}{(x_{23}^2)^\delta} \right]
\frac{1}{(x_{34}^2)^\delta}\left[ \hat \partial^{s-k}_1  \frac{1}{(x_{41}^2)^\delta} \right] + (x_3 \leftrightarrow x_4) \, , \nonumber \\
&K_2 (x_i) = 16N C_\phi^4 \sum_{k,l=0}^s a_k a_l \left[ \hat \partial^k_1  \frac{1}{(x_{13}^2)^\delta} \right] \left[ \hat \partial^{s-l}_2 \frac{1}{(x_{32}^2)^\delta} \right]
\left[ \hat \partial_2 ^{l} \frac{1}{(x_{24}^2)^\delta} \right] \left[ \hat \partial^{s-k}_1  \frac{1}{(x_{41}^2)^\delta} \right] \, .
\end{align}
Accordingly, we separate the integral $I_1$ into two parts as
\begin{align}
I_1 = I_1^{(1)} + I_1^{(2)} \, , \quad
 I_1^{(a)} =  \frac12 \int d^d x_3 d^d x_4 G(x_{34}) K_a (x_i)
\end{align}
with $a=1,2$. We start from the integral $I_1^{(1)}$ since it is the simplest one. We then evaluate the other integral $I_1^{(2)}$.

\subsubsection{Integral $I_1^{(1)}$}

We would like to compute the residue of
\begin{align}
I_1^{(1)} =&  16 N C_\phi^4 C_\sigma \sum_{k,l=0}^s a_k a_l  \left[ \hat \partial^k_1 \hat \partial^{s-l}_2 \frac{1}{(x_{12}^2)^\delta} \right]    \hat \partial^l_2 \hat \partial^{s-k}_1 
\int d^d x_3 d^d x_4  \frac{1}{(x_{23}^2)^{\delta} (x_{34}^2)^{\delta + 2 - \Delta} (x_{41}^2)^\delta}  
\end{align}
at $\Delta = 0$. 
The integral over $x_3$ and $x_4$ can be performed by applying \eqref{chain} as
\begin{align}
\nonumber
\int d^d x_3 d^d x_4  \frac{1}{(x_{23}^2)^{\delta} (x_{34}^2)^{\delta + 2 - \Delta} (x_{41}^2)^\delta}  &= v(\delta , \delta + 2 - \Delta , \Delta) v(\delta + 1 - \Delta , \delta , 1 + \Delta) \frac{1}{(x_{12}^2)^{\delta - \Delta}} \\
& = - \frac{1}{\Delta} \frac{\pi^d}{\Gamma(\delta + 2) \Gamma(\delta)}  \frac{1}{(x_{12}^2)^{\delta}} + \mathcal{O} (\Delta^0)\, .  
\end{align}
Thus we have 
\begin{align}
 \left( I_1^{(1)} \right)^\text{res} = - \frac{8 C_\phi^2 C_\sigma \pi^d}{\Gamma(\delta + 2) \Gamma (\delta)} D_0^s   = - 2 \gamma_\phi D_0 ^s \, ,
 \label{resI11s}
\end{align}
where we have used \eqref{current2pt} and \eqref{Csigma}.

\subsubsection{Integral $I_1^{(2)}$}
\label{I12}

\begin{figure}
 \centering
 \includegraphics[keepaspectratio, scale=0.45]
      {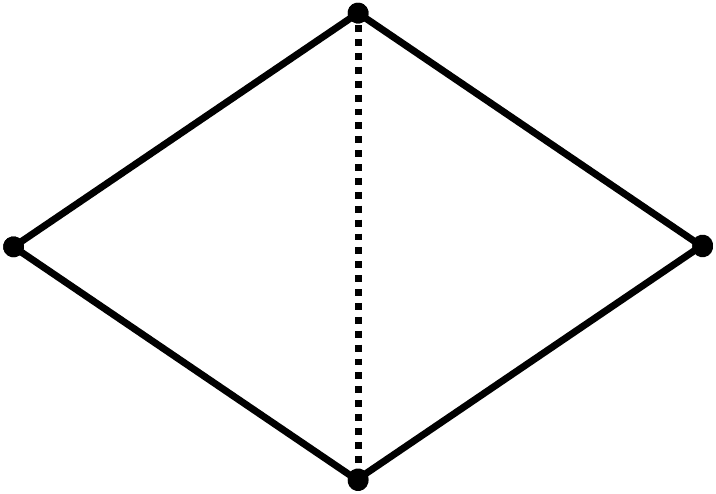}
 \caption{The graphical expression of integral \eqref{Bkl}. The solid line and the dotted line represents propagators with dimension $\delta$ and $2 - \Delta$, respectively. Derivatives may be acted on the solid lines.}
 \label{2loop}
\end{figure}

We then examine the integral
\begin{align}
 I_1^{(2)} =  &  8 N C_\phi^4 C_\sigma \sum_{k,l=0}^s a_k a_l  B_{k,l}
 \end{align}
 with
 \begin{align}
B_{k,l} =  \int d^d x_3 d^d x_4  \left[ \hat \partial^k_1  \frac{1}{(x_{13}^2)^\delta} \right] \left[ \hat \partial^{s-l}_2 \frac{1}{(x_{32}^2)^\delta} \right] \left[ \hat \partial^{s-k}_1  \frac{1}{(x_{41}^2)^\delta} \right]
 \left[ \hat \partial_2 ^{l} \frac{1}{(x_{24}^2)^\delta} \right]  \frac{1}{(x_{34}^2)^{2 - \Delta}}\, . \label{Bkl}
\end{align}
The integral can be expressed graphically as in fig.~\ref{2loop}.
It is argued to be possible to perform the integration in the coordinate representation  \cite{Vasiliev:1981dg}. However, for our purpose, it is suitable to work in the momentum representation. Applying the Fourier transformation formula \eqref{Fourier},
 the integral can be rewritten as
 \begin{align}
&B_{k,l} = \frac{C_B}{(2 \pi)^{4d}} 
 \int d^d x_3 d^d x_4 \prod_{n=1}^5 d ^dp_n  \frac{(i \hat p_1)^k (i \hat p_2)^{s-l} (i \hat p_3)^{s-k}  (i \hat p_4)^{l} }{p_1^2 p_2^2p_3^2p_4^2 (p_5^2)^{d/2 - 2 + \Delta}} 
 e^{i (p_1 \cdot x_{13} + p_2 \cdot x_{23} + p_3 \cdot x_{14} + p_4 \cdot x_{24} + p_5 \cdot x_{34} ) }
 \nonumber
\end{align}
with $\hat p_n = \epsilon \cdot p_n$ and
\begin{align}
C_B = ( C_\phi)^{-4}  C_{2 - \Delta} \, , \quad 
C_{2 - \Delta} = \frac{a(2 - \Delta)}{\pi^{d/2} 2^{4 - 2 \Delta}} \, .
\end{align}
Integration over $x_3$ and $x_4$ leads to the delta functions as
$(2\pi)^{2d} \delta^{(d)}(p_1 + p_2 - p_5) \delta^{(d)}(p_3 + p_4 + p_5)$,
and we eliminate $p_2$ and $p_5$ by using them.
We change the rest of integral variables as
\begin{align}
p_1 = p -q \, , \quad p_3 = q \, , \quad p_4 = - q'  \, .
\end{align}
This also means $p_2 =  q' - p$ and $p_5 =q' - q $. 
With these variables, the integral is now
 \begin{align}
&B_{k,l} = \frac{C_B}{(2 \pi)^{2d}}
 \int d^d p d^d q d^d q'
 \frac{(i (\hat p - \hat q) )^k (i (\hat q' - \hat p ))^{s-l} (i \hat q)^{s-k}  (-i \hat q')^{l} }{(p-q)^2 (p - q')^2 q^2 {q'}^2 ((q -q ')^2)^{d/2 - 2 + \Delta}} 
 e^{i p \cdot x_{12} } \, .
 \nonumber
\end{align}
If we think $p$ as an external momentum, then the integral is two-loop one with two internal momenta $q$ and $q'$. There are two ways to have divergence by collapsing the right loop first or the left one first, see fig.~\ref{2loop}. Therefore, we have to sum up two contributions obtained by integrating $q$ first and by integrating $q'$ first. Both contributions are the same, so we  multiply the result the factor 2 at the end of the computation.

We first evaluate the $q$-integral as
\begin{align}
\int \frac{d^d q}{(2 \pi)^d}  \frac{(i (\hat p - \hat q) )^k  (i \hat q)^{s-k}   }{q^2  (p-q)^2  ((q -q ')^2)^{d/2 - 2 + \Delta}} \, .
\end{align}
Introducing the Feynman parameters $x,y$ we arrive at
\begin{align}
\nonumber
&\frac{\Gamma (d/2 + \Delta)}{\Gamma(d/2 - 2 + \Delta)}  \int_0^1 dx \int_0^{1-x} dy \int
\frac{d^d q}{(2 \pi)^d}
 \\ & \qquad  \qquad \qquad \times   \frac{(i (\hat p - \hat q) )^k  (i \hat q)^{s-k}   y^{d/2 - 3 + \Delta}}
{((q - x p - y q')^2 + x(1-x) p^2 + y(1 - y) {q'}^2  - 2 xy p \cdot q')^{d/2 + \Delta} } \nonumber \\
& = \frac{\Gamma (d/2 + \Delta)}{\Gamma(d/2 - 2 + \Delta)} \int_0^1 dx \int_0^{1-x} dy \int
\frac{d^d w}{(2 \pi)^d}  
 \\ & \qquad \qquad \qquad  \times 
  \frac{(i (- \hat w + (1- x) \hat p - y \hat q') )^k  (i (\hat w + x \hat p + y \hat q' ))^{s-k}  y^{d/2 - 3 + \Delta}}
{(w^2 + x(1-x) p^2 + y(1 - y) {q'}^2 - 2 xy p \cdot q' )^{d/2 + \Delta} } \nonumber \\
&= \frac{1}{(4 \pi)^{d/2}} \frac{\Gamma(\Delta)}{\Gamma (d/2-2 +\Delta)}
\int_0^1 dx \int_0^{1-x} dy  \frac{(i ( (1- x) \hat p - y \hat q') )^k  (i ( x \hat p + y \hat q' ))^{s-k}  y^{d/2 - 3 + \Delta}}
{( x(1-x) p^2 + y(1 - y) {q'}^2 - 2 xy p\cdot q' )^{ \Delta} } \, .  \nonumber
\end{align} 
In the second equality, we have used the fact that the integral with $(\epsilon \cdot w)^n$ in the numerator vanishes for non-zero $n$. 
For odd $n$, we can see this from the symmetry argument.
For even $n$, the integral with $w^{\mu_1} \cdots w^{\mu_n}$ is proportional to the $n/2-$th power of $g^{\mu_i \mu_j}$. After contracting with $\epsilon^{\mu_i}$, we have  $(\epsilon \cdot \epsilon)^{n/2} = 0$ except for $n=0$. 
Since we are interested in $1/\Delta$-pole, we pick up the singular part as
\begin{align}
 \frac{1}{\Delta} \frac{1}{(4 \pi)^{d/2}} \frac{1}{\Gamma (d/2-2 )}
\int_0^1 dx \int_0^{1-x} dy  (i ( (1- x) \hat p - y \hat q') )^k  (i ( x \hat p + y \hat q' ))^{s-k}   y^{d/2 - 3 }  \, .
\end{align}
The integrals over $x$ and $y$ are
\begin{align}
&(i)^{s} \sum_{a = 0}^k \sum_{b = 0}^{s-k} \binom{k}{a} \binom{s-k}{b} 
\int_0^1 dx \int_0^{1-x} dy (1-x)^{k - a} x^{s-k-b} (-1)^a y^{a + b+ d/2 - 3} {\hat p}^{s - a - b}  \hat {q '}^{a + b} \nonumber \\
&=(i)^{s} \sum_{a = 0}^k \sum_{b = 0}^{s-k} \binom{k}{a} \binom{s-k}{b}  \frac{(-1)^a{\hat p}^{s - a - b}  \hat {q '}^{a + b} }{a + b+ d/2 - 2} 
\int_0^1 dx   (1-x)^{k + b+ d/2 - 2} x^{s-k-b}   \\
& =(i)^{s} \sum_{a = 0}^k \sum_{b = 0}^{s-k} \binom{k}{a} \binom{s-k}{b}  \frac{(-1)^a{\hat p}^{s - a - b}  \hat {q '}^{a + b} }{a + b+ d/2 - 2} 
\frac{\Gamma(k + b + d/2 - 1) \Gamma (s-k - b + 1)}{\Gamma (s + d/2)} \, . \nonumber
\end{align}
The $q'$-integral can be evaluated similarly as
\begin{align}
&(-i)^{s}
\int \frac{d^d q'}{(2 \pi)^d} \frac{(\hat p - \hat q ')^{s-l} (\hat q' )^{l + a + b}}{(p-q')^2{q '}^2} \\
&= (-i)^{s} \frac{\Gamma(2 - d/2)}{(4 \pi)^{d/2}} 
 \int_0^1 dz \frac{((1-z )\hat p)^{s-l} (z \hat p)^{l + a + b}}{(z(1-z) p^2)^{2 - d/2}} \nonumber \\
 &= (-i)^{s} \frac{\Gamma(2 - d/2)}{(4 \pi)^{d/2}} \frac{\Gamma (s - l - 1 + d/2) \Gamma (l + a + b - 1 + d/2)}{\Gamma (s + a + b - 2 + d)} \frac{\hat p^{s + a + b}}{(p^2)^{2 - d/2}} \, .\nonumber
\end{align}
The $p$-integral is
\begin{align}
\label{pint}
 \int d^d p \frac{\hat p^{2s}}{(p^2)^{2- d/2}} e^{i p \cdot  x_{12}}
& = (-i)^{2s} \frac{\pi^{d/2} 2^{2d -4} \Gamma(d-2)}{\Gamma (2 - d/2)} 
 2^{2s} \frac{\Gamma (2s + 2\delta)}{\Gamma (2\delta)} \frac{(\hat x_{12})^{2s}}{(x_{12}^2)^{2 \delta + 2s}} \\
 &= (-1)^{s} \frac{\pi^{3 d/2} 2^{2d +1}  (2s + 2\delta - 1)  \Gamma(2 \delta - 1 + s)}{\Gamma (2 - d/2) (\Gamma (\delta))^2 N \Gamma (s+1)} 
  D_0^s\, . \nonumber
\end{align}

Combining the above computations and multiplying the factor 2, we obtain
\begin{align}
B_{k,l} = \frac{1}{\Delta}  
  \frac{  (2s + d - 3 )  \Gamma(d-3 + s)}{4 N C_\phi^4 \Gamma(s+1)(\Gamma(\delta))^2}  D_0^sH_1^{(2)} (k,l) + \mathcal{O}(\Delta^0)
  \label{Bklsum}
\end{align}
with
\begin{align}
\label{H12}
& H_1^{(2)} (k,l) = \sum_{a=0}^k \sum_{b=0}^{s-k} \binom{k}{a} \binom{s-k}{b}\frac{(-1)^{s + a}}{a + b + d/2 - 2}
 \\
 & \quad  \times   \frac{\Gamma (k + b + d/2 - 1) \Gamma (s - k - b + 1)}{\Gamma (s + d/2)} \frac{\Gamma (s-l - 1 + d/2) \Gamma (  l + a + b - 1 + d/2)}{\Gamma (s + a + b - 2 +d)} \, . \nonumber
\end{align}
Thus we find
\begin{align}
I_1^{(2)} = \frac{1}{\Delta} 
  \frac{2  (2s + d - 3 )  \Gamma(d-3 + s)}{\Gamma(s+1) (\Gamma(\delta))^2}  C_\sigma D_0^s
   \sum_{k,l=0}^s a_k a_l H_1^{(2)} (k,l) + \mathcal{O}(\Delta^0) \, .
\end{align}
After summing over $k,l,a,b$,%
\footnote{\label{Mathematica}%
We have checked the identity for explicit values of spin $s=1,2,\cdots$ and real dimension $d$ by  Mathematica. 
The same is true for the other expressions after the summations in \eqref{resI2}, \eqref{resI12f}, \eqref{resI2f}, \eqref{resa0} and \eqref{altI12}.
 It might be possible to proof these identities analytically by applying the method in \cite{Diaz:2006nm}, see also appendix \ref{app2pt}.} 
we find
\begin{align}
 \left( I_1^{(2)} \right)^\text{res}=  \frac{d (d-2) }{2 (s + d/2 - 2) (s + d/2 - 1)} \gamma_\phi D_0^s \, .
\label{resI12}
\end{align}
Combining \eqref{resI11s}, we thus find the residue of $I_1$ at $\Delta = 0$
as in \eqref{scalarresult}.

\subsection{Integral $I_2$}
\label{I2}

\begin{figure}
 \centering
 \includegraphics[keepaspectratio, scale=0.5]
      {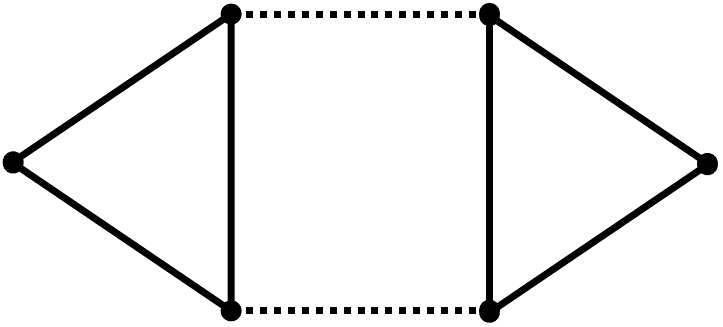}
 \caption{The graphical expression of the integral in \eqref{Ckl} with possible derivatives on the solid lines. }
 \label{3loop}
\end{figure}

We move to the computation of  $I_2$.
With the Wick contraction \eqref{Wicks}, the three point function is computed as
\begin{align}
\langle J_s^\epsilon (x_1) \mathcal{O} (x_3) \mathcal{O} (x_4) \rangle_0
= 8N C_\phi^3 \frac{1}{(x_{34}^2)^\delta} \sum_{k=0}^s
  a_k \left[ \hat \partial^{s-k}_1 \frac{1}{(x_{41}^2)^\delta} \right]
  \left[ \hat \partial^{k}_1 \frac{1}{(x_{13}^2)^\delta} \right] \, .
  \label{threepts}
\end{align}
Thus the integral $I_2$ is written as
\begin{align}
&I_2 = 2^5 N^2 C_\phi^6 C_\sigma^2 \sum_{k,l=0}^s a_k a_l C_{k,l} 
\end{align}
with
\begin{align}
\label{Ckl}
&C_{k,l} = 
\int d^d x_3  d^d x_4 d^d x_5 d^d x_6  \\
&\times \left[ \hat \partial^{s-k}_1 \frac{1}{(x_{41}^2)^\delta} \right]
  \left[ \hat \partial^{k}_1 \frac{1}{(x_{13}^2)^\delta} \right] \left[ \hat \partial^{s-l}_2 \frac{1}{(x_{62}^2)^\delta} \right]
    \left[ \hat \partial^{l}_2 \frac{1}{(x_{25}^2)^\delta} \right] \frac{1}{(x_{34}^2)^\delta} \frac{1}{(x_{56}^2)^\delta} \frac{1}{(x_{35}^2)^{2 - \Delta}} \frac{1}{(x_{46}^2)^{2 - \Delta}}\, . \nonumber
\end{align}
The integral can be represented graphically as in  fig.~\ref{3loop}.
Without derivatives, this integral is evaluated to the order $\Delta^0$ in \cite{Vasiliev:1981dg} as
\begin{align}
 \left. C_{0,0} \right|_{s=0} = \frac{\pi^{2d} (a(\delta))^3 (a (2))^3 a (d-3)}{\Gamma(d/2)} \left(  \frac{1}{\Delta} + 4 B(2) - 3 B(\delta) - B(d - 3) \right)  \, ,
 \label{vasiliev}
\end{align}
where $B(x) = \psi (x) + \psi (d/2 - x)$  with $\psi (x) = \frac{d}{dx} \log \Gamma (x)$.
Thus the task is to incorporate the effects of derivatives.

The $1/\Delta$-pole terms arise from sub-diagrams, and there are diagrams which 
reproduce the residue at $\Delta = 0$ (see fig.~\ref{3loop_0} for a related computation). In \cite{Vasiliev:1981dg}, they compute the order $\Delta^0$ contributions from the difference between the diagrams.
Since we are only interested in the singular terms, we may use the same diagrams  in  \cite{Vasiliev:1981dg}. However, the diagrams have different structure from the one in  fig.~\ref{3loop}, and it turns out to give wrong answers if derivatives are acted on lines. 
Therefore, we would like to utilize a diagram which has the same structure.

We propose to use a diagram where the exponent of the propagator is shifted of order $\Delta$ such that we can integrate over, say, $x_3,x_4$ by applying the formula \eqref{unique}. 
The amount of residue would change, so we will fix it by comparing the result without derivatives to the $1/\Delta$-pole term in \eqref{vasiliev}.
 Note that the procedure works only when the effects of derivatives do not give rise to any extra zeros nor poles. 
There are many ways to do this but we  shift the exponent of propagator between $x_3$ and $x_4$ as $\delta \to \delta +\Delta$ as a convenient choice.
After the integration, we would arrive at  essentially the same diagram in the previous section, so we can evaluate the residue in the momentum representation. 
As a confirmation of procedure, we re-derive \eqref{resI12} by applying this method in appendix \ref{alt}

Following the above argument, we compute the following integral as 
\begin{align}
\label{tCkl}
&C_{k,l} ' = 
\int d^d x_3  d^d x_4 d^d x_5 d^d x_6  \\
&\times \left[ \hat \partial^{s-k}_1 \frac{1}{(x_{41}^2)^\delta} \right]
  \left[ \hat \partial^{k}_1 \frac{1}{(x_{13}^2)^\delta} \right] \left[ \hat \partial^{s-l}_2 \frac{1}{(x_{62}^2)^\delta} \right]
    \left[ \hat \partial^{l}_2 \frac{1}{(x_{25}^2)^\delta} \right] \frac{1}{(x_{34}^2)^{\delta + \Delta}} \frac{1}{(x_{56}^2)^\delta} \frac{1}{(x_{35}^2)^{2 - \Delta}} \frac{1}{(x_{46}^2)^{2 - \Delta}} \, .\nonumber
\end{align}
Using \eqref{unique} the integral over $x_3$ can be evaluated as
\begin{align}
& \hat \partial^{k}_1  \int d^d x_3 
   \frac{1}{(x_{13}^2)^\delta (x_{34}^2)^{\delta + \Delta} (x_{35}^2)^{2 - \Delta}}  = v(\delta , \delta + \Delta , 2 - \Delta)
  \hat \partial^k_1 \left[ \frac{1}{x_{45}^2 (x_{15}^2)^{1 - \Delta} (x_{14}^2)^{d/2 - 2 +\Delta} } \right] \nonumber \\
 & =  \frac{\pi^{d/2} \Gamma(1 - \Delta) \Gamma(d/2-2 + \Delta)}{\Gamma(\delta) \Gamma(\delta + \Delta) \Gamma(2 - \Delta)} \frac{1}{x_{45}^2} 
 \sum_{a=0}^k \binom{k}{a}  \left[ \hat \partial^a_1  \frac{1}{(x_{15}^2)^{1 - \Delta}} \right] \left[ \hat \partial^{k-a}_1  \frac{1}{(x_{14}^2)^{d/2 - 2 +\Delta} } \right] \, .
\end{align}
The propagator between $x_1$ and $x_4$ can be rewritten as
\begin{align}
\left[ \hat \partial^{s-k}_1 \frac{1}{(x_{14}^2)^\delta} \right]
\left[ \hat \partial^{k-a}_1  \frac{1}{(x_{14}^2)^{d/2 - 2 +\Delta} } \right] 
= \frac{(\delta)_{s-k} (d/2 - 2 + \Delta)_{k-a}}{(d-3+\Delta)_{s-a}} 
\hat \partial^{s-a}_1 \frac{1}{(x_{14}^2)^{d - 3 + \Delta}} \, . 
\end{align}
Integration over $x_4$ then gives
\begin{align}
&\hat \partial^{s-a}_1 \int d^d x_4  \frac{1}{(x_{14}^2)^{d - 3 + \Delta} x_{45}^2 (x_{46}^2)^{2 - \Delta}}  \\
&=  v(d- 3 + \Delta , 1 , 2 - \Delta) \hat \partial^{s-a}_1 \left[ \frac{1}{(x_{56}^2)^{3 - d/2 -\Delta} (x_{16}^2 )^{\delta} (x_{15}^2)^{d/2 - 2 + \Delta} } \right] \nonumber \\
&=  \frac{\pi^{d/2} \Gamma(3 - d/2 - \Delta) \Gamma(\delta) \Gamma(d/2 - 2 + \Delta)}{\Gamma(d - 3 + \Delta ) \Gamma(2 -\Delta)} \frac{1}{(x_{56}^2)^{3 - d/2 -\Delta}} 
\nonumber \\ & \qquad \qquad \times  \sum_{b=0}^{s-a} \binom{s-a}{b} \left[ \hat \partial^{b}_1 
\frac{1}{ (x_{16}^2 )^{\delta}} \right] \left[ \hat \partial^{s-a-b}_1  \frac{1}{ (x_{15}^2)^{d/2 - 2 + \Delta} } \right] \nonumber \, .
\end{align}
We re-express the propagator between $x_1$ and $x_5$  as
\begin{align}
&\left[ \hat \partial^a_1  \frac{1}{(x_{15}^2)^{1 - \Delta}} \right]
\left[ \hat \partial^{s-a-b}_1  \frac{1}{ (x_{15}^2)^{d/2 - 2 + \Delta} } \right] 
&= \frac{(1-\Delta)_a (d/2 -2 + \Delta)_{s-a-b}}{(\delta)_{s-b}}
\hat \partial^{s-b}_1  \frac{1}{ (x_{15}^2)^{\delta} } \, . 
\end{align}
The integral over $x_5$ and $x_6$ is actually the same as $B_{b,l}$ defined in \eqref{Bkl} and evaluated as in \eqref{Bklsum} with \eqref{H12}.

Summing over all contributions, we find
\begin{align}
& C_{k,l} ' = \frac{1}{\Delta} \frac{(2s + d - 3) \Gamma(d - 3 + s ) \pi^d \Gamma(3 - d/2)}{4 N C_\phi^4 \Gamma(s+1) (\Gamma (\delta))^3} D_0^s H_2(k,l)   + \mathcal{O} (\Delta^0) \, ,
\end{align}
where
\begin{align}
 H_2 (k,l) =& \sum_{a=0}^k \sum_{b=0}^{s-a} \binom{k}{a} \binom{s-a}{b}  \\ & \times
 \frac{\Gamma(s - k + \delta) \Gamma (k - a + d/2 - 2 ) \Gamma (a+1) \Gamma (s- a - b + d/2 - 2)}{\Gamma (s - a + d - 3) \Gamma (s - b + \delta)} H_1^{(2)} (b ,l) 
 \, . \nonumber
\end{align}
We can see that there is no extra zero or pole at $\Delta = 0$ for generic $d$, so we can confirm that the effect of derivatives is the same between $C_{k,l}$ and $C_{k,l} ' $.
However, there is no guarantee that the overall factor is the same as mentioned above. In fact, by comparing $C_{0,0} '$ at $s=0$ with the $1/\Delta$-pole term in \eqref{vasiliev}, we find
\begin{align}
C_{k,l} = \frac{1}{2} C_{k,l} ' + \mathcal{O}(\Delta^0) \, .
\end{align}
The integral $I_2$ thus becomes
\begin{align}
\label{resI2}
I_2 &=  2^5 N^2 C_\phi^6 C_\sigma^2 \sum_{k,l =0}^s a_k a_l C_{k,l}  \\
& = \frac{1}{\Delta} \frac{2 (2s + d - 3) \Gamma(d - 3 + s ) \Gamma(d+1) \gamma_\phi}{\Gamma(s+1) (\Gamma (\delta))^2( \Gamma(\delta - 1))^2 (d-1)}D_0^s \sum_{k,l =0}^s a_k a_l  H_2(k,l)  + \mathcal{O} (\Delta^0) \, . \nonumber
\end{align}
The summation over $k,l$ can be performed as in \eqref{scalarresult}.

\section{Gross-Neveu model}
\label{GN}

In the previous sections, we have developed the formalism to compute the anomalous dimensions of higher spin current in the framework of conformal perturbation theory. In this section, we apply the methods to the Gross-Neveu model in  $d$ dimensions as well. We will see that the direct application does not work due to an extra divergence which cannot be resolved by introducing $\Delta$ in \eqref{Greg}. In order to regularize the divergence, we introduce another shift of exponent by $\eta/2$ in the free fermion propagator. The final result turns out to be finite in the limit $\eta \to 0$.
After the slight modification, we succeed to reproduce the anomalous dimensions previously obtained in \cite{Muta:1976js}.

\subsection{Anomalous dimensions of higher spin currents}

We consider the system of $\tilde N$ Dirac fermions $\psi^i$ $(\bar \psi^i)$ with $i = 1,2, \cdots , \tilde N$,
which transform in the (anti-)fundamental representation under the action of global U$(\tilde N)$ symmetry.
According to \cite{Anselmi:1999bb},
 the current of spin $s=1,2,\cdots$ can be written in terms of bi-linears of free fermions as%
 \footnote{The expression of currents in terms of Gegenbauer polynomial can be found in \cite{Skvortsov:2015pea}.}
\begin{align}
J_s^\epsilon = \sum_{k=0}^{s-1} \tilde a_k \hat \partial^k \bar \psi ^i \hat \gamma \hat \partial ^{s-k-1} \psi _i \, ,
\label{hscurrent}
\end{align}
where
\begin{align}
\tilde a_k = (-1)^k  \frac{\binom{s-1}{k}\binom{s+d-3}{k+d/2 - 1}}{\binom{s+d-3}{d/2-1}}  \, ,
\quad  \tilde a_{s-1-k} = (-1)^{s-1} \tilde a_k \, .
\label{tildeak}
\end{align}
We also use $\hat \gamma \equiv  \epsilon \cdot \gamma = \slashed{\epsilon}$.

We compute the two point function of the higher spin current
\begin{align}
\tilde D_0^s \equiv \langle J_s^\epsilon (x_1  ) J_s^\epsilon (x_2 ) \rangle = \tilde C_s \frac{( \hat x_{12})^{2s}}{(x_{12}^2)^{2 \delta + 2s}} 
\label{2pt}
\end{align}
by applying  the Wick contraction as
\begin{align}
\label{Wickf}
\langle \psi ^i (x_1) \bar \psi ^j (x_2) \rangle = C_\phi \slashed{\partial}_1 \frac{\delta^{ij}}{(x_{12}^2)^{\delta}} \, , \quad  C_\phi = \frac{\Gamma(\delta)}{4 \pi^{d/2}} \, .
\end{align}
Using \eqref{hscurrent} the  two point function can be rewritten as
\begin{align}
\tilde D_0^s&= - C_\phi^2 \tilde N \sum_{k,l = 0}^{s-1}
\tilde a_k \tilde a_l g_4
   \left( \epsilon^{\mu_1}  \hat \partial_1^{s - k - 1} \hat \partial_2^l \partial_1^{\mu_2} \frac{1}{(x^2_{12})^\delta} \epsilon^{\mu_3} \hat \partial_2^{s - l - 1} \hat \partial_1^k \partial^{\mu_4}_2 
   \frac{1}{(x_{21}^2)^\delta}
     \right) \, .
\end{align}
Here we have set 
\begin{align}
\label{gammaab}
g_{n} = \text{tr} (\gamma_{\mu_1} \cdots \gamma_{\mu_n}) \, .
\end{align}
The factor $(-1)$ comes from the fermion loop. For the trace of gamma matrix we use
\begin{align} 
g_4 = 
  \text{tr} (\gamma_{\mu_1} \gamma_{ \mu_2} \gamma_{ \mu_3 } \gamma_{ \mu_4})
   = \text{tr} \mathbbm{1} (\delta_{\mu_1 \mu_2} \delta_{\mu_3 \mu_4} - \delta_{\mu_1 \mu_3} \delta_{\mu_2 \mu_4} + \delta_{\mu_1 \mu_4} \delta_{\mu_2 \mu_3}) \, .
   \label{gammatrace}
\end{align}
Using $\epsilon \cdot \epsilon = 0$, the above expression reduces to
\begin{align}
\nonumber \tilde D_0^s 
 &= (-1)^{s+1} 2 N C_\phi^2 \sum_{k,l = 0}^{s-1}
\tilde a_k \tilde a_l 
   \left(  \hat \partial_2^{s - k + l}  \frac{1}{(x_{12}^2)^\delta}  \hat \partial_2^{s +k - l } 
   \frac{1}{(x_{21}^2)^\delta}
     \right) \\
      \quad& = (-1)^{s+1} 2 N C_\phi^2 \sum_{k,l = 0}^{s-1}
     \tilde a_k \tilde a_l (\delta)_{s-k+l} (\delta)_{s+k-l} \frac{(2 \hat x_{12})^{2s}}{(x_{12}^2)^{ 2 \delta + 2 s}} \, . \label{tildeD}
\end{align}
Here we have set $N \equiv  \tilde N \text{tr} \mathbbm{1}$.
Thus we obtain $\tilde C_s$ in \eqref{2pt} as
\begin{align}
\tilde C_s =  2 ^{2s+1} N C_\phi^2 \Gamma (s)  \delta^2 (2 \delta  + s)_{s-1}  \, ,
 \label{2ptn}
\end{align}
where we have used 
\begin{align}
 \sum_{k,l = 0}^{s-1} (-1)^{s+1}
     \tilde a_k \tilde a_l (\delta)_{s-k+l} (\delta)_{s+k-l}
  =\Gamma (s) \delta^2  (2 \delta+s)_{s-1} 
  \label{formula1}
\end{align}
as shown in appendix \eqref{app2pt}.

We are interested in anomalous dimensions of higher spin currents after the deformation of \eqref{def}.
In the current case, the scalar operator is
\begin{align}
{\mathcal{O}}_{d-1} (x) \equiv
\tilde{\mathcal{O}} (x) = \bar \psi^i \psi_i \, ,
\label{GNso}
\end{align}
whose normalization is
\begin{align}
\langle \tilde{\mathcal{O}}(x_1) \tilde{\mathcal{O}} (x_2) \rangle
 = - \tilde N C_\phi^2 \text{tr} \left ( \slashed{\partial}_1 \frac{1}{(x_{12}^2)^\delta} \slashed{\partial}_2 \frac{1}{(x_{21}^2)^\delta}\right)  = C_{\tilde{\mathcal{O}}}\frac{1}{(x_{12}^2)^{d-1} }\, , \quad
C_{\tilde{\mathcal{O}}} = 4 N \delta^2 C_\phi^2 \, .
\end{align}
Here we have used $\text{tr} (\gamma_\mu \gamma_\nu) = \text{tr} \mathbbm{1} \delta_{\mu \nu}$. 
The operator is dual to the scalar field in the type B Vasiliev theory, and the deformation changes the boundary condition for the bulk scalar from the Dirichlet one at $f=0$ to the Neumann one at $f=\infty$.
The propagator \eqref{Greg} becomes
\begin{align}
G_{d-1} (x_{12}) \equiv \tilde G (x_{12}) = C_{\tilde \sigma}\frac{1}{(x_{12}^2)^{1 - \Delta}} \, , \quad 
C_{\tilde \sigma} = \frac{(1 - d/2) \sin (\pi d /2) \Gamma(d-1)}{C_{\tilde{ \mathcal{O}}} \pi^{d+1}  }  \, .
\end{align}
It might be convenient to express the coefficient $C_{\tilde \sigma}$ as%
\footnote{See, e.g.,  \cite{Gracey:1990sx,Gracey:1991vy,Gracey:1990wi,Derkachov:1993uw,Vasiliev:1996rd,Vasiliev:1997sk} for previous studies on the Gross-Neveu model in $d=2 + \epsilon$ dimensions.}
\begin{align}
C_{\tilde \sigma} =  2 d \gamma_\psi \, , \quad
\gamma_\psi = \frac{\delta^2 \Gamma(d -1)}{N\Gamma (2 - d/2) \Gamma(d/2 + 1)\Gamma(d/2)^2} 
\label{anopsi}
\end{align}
in terms of the anomalous dimension of $\psi^i$ to the $1/N$ order.

In the following we shall compute the residues of the integrals $I_1$ and $I_2$ in \eqref{integrals} at $\Delta = 0$ by following the analysis in the previous section with a slight modification.
The results are summarized as
\begin{align}
&I_1^\text{res} = - 2 \gamma_\phi \left(1  -   \frac{d (d-2) }{4 (s + d/2 - 2) (s + d/2 - 1)} \right) \tilde D_0^s 
 = - 2 \gamma_\psi \frac{(s-1)(s + d - 2) }{ (s + d/2 - 2) (s + d/2 - 1)} \tilde D_0^s  \, , \nonumber \\
&I_2^\text{res} = \gamma_\psi \frac{P_s \Gamma(d+1) \Gamma( s + 1) }{ 2 (d-1) \Gamma (d + s - 3) ( s + d/2 - 2) (s + d/2 - 1)} \tilde D_0^s  \, ,
\label{fermionresult}
\end{align}
where
\begin{align}
P_s = \frac{1+(-1)^s}{2} \, .
\label{Projector}
\end{align}
These integrals have been evaluated in \cite{Diab:2016spb} for spin $s=1,2$ up to $\mathcal{O}(\Delta^0) $ terms, and we find agreement. For spin $s \geq 3$ our results are new.
 The anomalous dimension $ \gamma_s$ can be read off as
 \begin{align}
 \label{anomalousf}
\gamma_s = 2 \gamma_\psi \frac{1}{ (s + d/2 - 2) (s + d/2 - 1)} \left[ (s-1)(s + d - 2) - \frac{P_s  \Gamma(d+1) \Gamma( s + 1) }{ 2 (d-1) \Gamma (d + s - 3) }  \right] \, ,
 \end{align}
which reproduces the result in \cite{Muta:1976js}. 
We can check that the anomalous dimension vanishes for $s=2$, which implies the conservation of conformal symmetry. We can also see $\gamma_1 = 0$ for spin one current, which means that the global $\text{U}(1) (\in \text{U}(\tilde N))$ symmetry is not broken. 
The expression $\gamma_s$ in \eqref{anomalousf} coincides with that in \eqref{anomalouss} for the scalar case with even spin if we replace $\gamma_\psi$ by $\gamma_\phi$. 

\subsection{Integral $I_1$}

The integral $I_1$ involves four point function with two higher spin currents.
The four point function may be divided into two parts as
\begin{align}
\langle J_s^\epsilon (x_1 ) J_s^\epsilon (x_2 )  \tilde{\mathcal{O}}  (x_3) \tilde{\mathcal{O}} (x_4) \rangle _0
 = \tilde K_1 (x_i) + \tilde K_2 (x_i) \, ,
 \label{4ptp}
\end{align}
where the two functions $\tilde K_a (x_i)$ $(a=1,2) $ are 
\begin{align}
\nonumber
\tilde K_1 (x_i )
 &= - 2 \tilde N C_\phi^4 \sum_{k,l = 0}^{s-1} \tilde a_k \tilde a_l g_{6}
  \hat \partial^k _1 \partial_4^{\mu_1} \frac{1}{(x_{41}^2)^\delta} \epsilon^{\mu_2} 
  \hat \partial_1^{s-1-k} \hat \partial_2^l \partial_1^{\mu_3} \frac{1}{(x_{12}^2)^\delta} 
  \epsilon^{\mu_4} \hat \partial^{s-1-l}_2 \partial^{\mu_5}_2 \frac{1}{(x_{23}^2)^\delta}
  \partial_3^{\mu_6} \frac{1}{(x_{34}^2)^\delta}  
  \\& + (x_3 \leftrightarrow x_4) \, , 
 \label{4ptp1}
\end{align}
and 
\begin{align}
\nonumber
\tilde K_2 (x_i )
 &= - 2 \tilde N  C_\phi^4 \sum_{k,l = 0}^{s-1} \tilde a_k \tilde a_l  g_{6}
  \hat \partial^k _1 \partial_4^{\mu_1} \frac{1}{(x_{41}^2)^\delta} \epsilon^{\mu_2} 
  \hat \partial_1^{s-1-k} \partial_1^{\mu_3} \frac{1}{(x_{13}^2)^\delta} 
 \hat \partial^{l}_2 \partial^{\mu_4}_3 \frac{1}{(x_{32}^2)^\delta} \epsilon^{\mu_5} 
 \hat \partial^{s-1- l}_2 \partial_2^{\mu_6} \frac{1}{(x_{24}^2)^\delta}  \, .
 \label{4ptp2}
\end{align}
In terms of $\tilde K_a$, the integral $I_1$ in \eqref{integrals} becomes
\begin{align}
I_1 = I_1^{(1)} + I_1^{(2)} \, , \quad
I^{(a)}_1 = \frac{1}{2 }  \int d^d x_3 d^d x_4 
 \tilde K_a (x_i ) \tilde G(x_{34}) \, .
\end{align}
In the following subsections, we compute the residues of $I_1^{(1)}$ and $I_1^{(2)}$ at $\Delta = 0$. 

\subsubsection{Integral $I_1^{(1)}$}

We start from the simplest integral $I_1^{(1)}$.
Rewriting
\begin{align}
\left( \partial_3^{\mu_6} \frac{1}{(x_{34}^2)^\delta} \right) \frac{1}{(x_{34}^2)^{1 - \Delta}}
= \frac{d - 2}{d - 2 \Delta} \partial_3^{\mu_6} \frac{1}{(x_{34}^2)^{d/2 - \Delta}}
\end{align}
and integrating by part, we obtain
\begin{align}
I_1^{(1)} =& - 2 \tilde N C_{\tilde \sigma} C_\phi^4 \frac{d-2}{d - 2 \Delta}  \sum_{k,l = 0}^{s-1}\tilde a_k \tilde a_l g_{6} (-1)^l \hat \partial_{1}^{s - 1 - k +l} \epsilon^{\mu_2 } \epsilon^{\mu_4} \partial_1^{\mu_3} \frac{1}{(x_{12}^2)^\delta} \\
&\quad \times
\hat \partial_1^k \partial^{\mu_1}_1 \hat \partial^{s - 1 -l}_2 \partial_2^{\mu_5} \partial_2^{\mu_6} \int d^3 x_3 d^3 x_4 \frac{1}{(x_{41}^2)^\delta} \frac{1}{(x_{23}^2)^\delta} \frac{1}{(x_{34}^2)^{d /2 - \Delta}} \nonumber \\ 
=& - 4 N C_{\tilde \sigma} C_\phi^4 \frac{d-2}{d - 2 \Delta} \sum_{k,l = 0}^{s-1}\tilde  a_k \tilde a_l  (-1)^l \hat \partial_{1}^{s  - k +l}  \frac{1}{(x_{12}^2)^\delta} \nonumber \\
&\quad \times
\hat \partial_1^{k+1} \hat \partial^{s - 1 -l}_2 \square_2\int d^3 x_3 d^3 x_4 \frac{1}{(x_{41}^2)^\delta} \frac{1}{(x_{23}^2)^\delta} \frac{1}{(x_{34}^2)^{d/2 - \Delta}} \, . \nonumber
\end{align}
In the second equality, we have used
\begin{align}
\gamma_{\mu_5} \gamma_{\mu_6} \partial_{2}^{\mu_5} \partial_2^{\mu_6} 
 = g_{\mu_5 \mu_6} \partial_{2}^{\mu_5} \partial_2^{\mu_6} 
 \equiv \square_2 
\end{align}
and \eqref{gammatrace}.

The integral over $x_3$ and $x_4$ can be done by applying the formula \eqref{chain} as
\begin{align}
&\int d^d x_3 d^d x_4 \frac{1}{(x_{41}^2)^\delta (x_{23}^2)^\delta (x_{34}^2)^{d/2 - \Delta} } \\
 & \qquad = v( \delta, d/2 - \Delta , 1 + \Delta) v (\delta , d/2 -1 - \Delta , 2 + \Delta)
\frac{1}{(x_{12}^2)^{d/2 - 2 - \Delta}}  \nonumber \\
& \qquad = \frac{1}{\Delta} \pi^d \frac{\Gamma(d/2 - 2)}{(\Gamma(\delta))^2 \Gamma(d/2)} \frac{1}{(x_{12}^2)^{d/2 - 2}} + \mathcal{O} (\Delta^0)\, . \nonumber
\end{align}
Utilizing
\begin{align}
 \square_2 (x_{12}^2)^{2 - d/2} = \partial_{2 \mu} \partial_2^\mu (x_{12}^2)^{2 - d/2} 
 = \partial_{2 \mu} \left(- 2 (2 - d/2) (x_{12})^\mu  (x_{12}^2)^{1 - d/2} \right)
  = \frac{ 2(4-d) }{(x_{12}^2)^{\delta}} \, , 
\end{align}
we find
\begin{align}
I_1^{(1)} =& \frac{1}{\Delta}  8 N C_{\tilde \sigma} C_\phi^4 \pi^d \frac{d-2}{\Gamma(\delta) \Gamma(d/2+1)}  \sum_{k,l=0}^{s-1} \tilde a_k \tilde a_l  (-1)^{s+1}  (\delta)_{s-k+l}  (\delta)_{s+k-l}  \frac{(2 \hat x_{12})^{2s}} {(x_{12}^2)^{2 \delta + 2 s}} 
  + \mathcal{O} (\Delta^0) \nonumber \\
  =& - \frac{1}{\Delta}  2 \gamma_\psi \tilde D_0^s 
    + \mathcal{O} (\Delta^0)  \, .
    \label{resI11f}
\end{align}
Here we have also used \eqref{tildeD}.

\subsubsection{Integral $I_1^{(2)}$}
\label{I12f}

As in the scalar case, we work in the momentum representation by applying the Fourier transform formula in \eqref{Fourier}.
Then the integral $I_1^{(2)}$ becomes
\begin{align}
I_1^{(2)} &= - \frac{\tilde  N C_{\tilde \sigma}  C_{1 - \Delta} }{(2 \pi)^{4d} }\int d^d x_3 d^d x_4 \prod_{n=1}^5 d ^d p_n
\sum_{k,l=0}^{s-1} \tilde a_k \tilde a_l \text{tr} (\slashed{p}_1 \slashed{\epsilon} \slashed{p}_2 \slashed{p}_3 \slashed{\epsilon} \slashed{p}_4)  \\ & \qquad  \times
\frac{ (i \hat p_1)^k (i \hat p_2)^{s - 1 - k} ( i \hat p_3)^l (i \hat p_4)^{s - 1 - l} 
 e^{i (p_1 \cdot x_{14} + p_2 \cdot x_{13} + p_3 \cdot x_{23} + p_4 \cdot x_{24} + p_5 \cdot x_{34})}} 
{p_1^2 p_2^2 p_3^2 p_4^2 (p_5^2)^{d/2 - 1 + \Delta}} \nonumber
\end{align}
with
\begin{align}
C_{1- \Delta} = \frac{a(1-\Delta) }{\pi^{d/2} 2^{2 - 2 \Delta}} \, .
\end{align}
Integration over $x_3$ and $x_4$ leads to delta functions as
$(2 \pi)^{2d} \delta^{(d)} (p_2 + p_3 - p_5) \delta^{(d)} (p_1 + p_4 + p_5)$.
Thus we may replace
\begin{align}
p_1 = p - q \, , \quad p_3 = - q' \, , \quad p_5 = q - q' 
\end{align}
with $p_2 = p_5 - p_3 = q$ and $p_4 = - p_1 - p_5 = q' - p$.
The integral is now given by
\begin{align}
\label{I12p}
I_1^{(2)} &=  \frac{ N C_{\tilde \sigma}  C_{1 - \Delta} }{(2 \pi)^{2d} } \sum_{k,l=0}^{s-1} \tilde a_k \tilde a_l \int d ^d p  \frac{J_{k,l} (\hat p)^{2s}}{(p^2)^{2 - d/2}} e^{i p \cdot x_{12}} \, , 
\end{align}
where $J_{k,l}$ are defined by
\begin{align}
\label{Jp}
 &J _{k,l}  \frac{(\hat p)^{2s}}{(p^2)^{2 - d/2 + \Delta}} \\
 & \equiv \int d ^d q d ^d q' 
\frac{1}{\text{tr}\mathbbm{1}}\text{tr} ((\slashed{p} - \slashed{q} ) \slashed{\epsilon} \slashed{q} \slashed{q}' \slashed{\epsilon} (\slashed{q} ' - \slashed{p}))  
\frac{ (\hat p - \hat q)^k ( \hat q)^{s - 1 - k} (  \hat q ' )^l  (\hat p - \hat q')^{s - 1 - l} 
 } 
{(p - q)^2 q^2 {q'}^2 (p - q' )^2 ((q - q')^2)^{d/2 -1 + \Delta}} \, . 
\nonumber
\end{align}

For the further computation, we  rewrite the trace over the product of gamma matrices in more convenient form.
Using \eqref{gammatrace} and $\epsilon \cdot \epsilon = 0$, we have 
\begin{align}
& \frac{1}{\text{tr}\mathbbm{1}} \text{tr} (\slashed{\epsilon} \slashed{q} \slashed{q}' \slashed{\epsilon} (\slashed{q} ' - \slashed{p}) (\slashed{p} - \slashed{q} ) )  =  \frac{2 \hat q}{\text{tr}\mathbbm{1}}\text{tr} ( \slashed{q}' \slashed{\epsilon} (\slashed{q} ' - \slashed{p}) (\slashed{p} - \slashed{q} ) )-  \frac{2 \hat q '}{\text{tr}\mathbbm{1}}  \text{tr}  ( \slashed{q} \slashed{\epsilon} (\slashed{q} ' - \slashed{p}) (\slashed{p} - \slashed{q} ) ) \nonumber \\
  &  = 2  (\hat q (\hat q ' - \hat p) q' \cdot (p - q) - \hat q (\hat p - \hat q) q' \cdot (q' - p) ) \nonumber
   \\  & \quad\quad  \quad\quad  +2  ( \hat q ' (\hat p - \hat q ) q \cdot (q' - p) - \hat q ' (\hat q' - \hat p) q \cdot (p - q) ) \nonumber 
\\
&= \hat q (\hat p - \hat q ' ) [ (p - q')^2- (q - q ')^2 - p^2 + q^2] \\
  & + \hat q (\hat p - \hat q) [p^2 - {q'}^2 - (p - q' )^2 ]+  \hat q ' (\hat p - \hat q ) [(p - q)^2 - (q - q ')^2 - p^2 + {q'}^2]  \nonumber \\
   & + \hat q ' (\hat p - \hat q' ) [p^2 - q^2- (p - q)^2 ]  ) \, . \nonumber 
\end{align}
Since $p-q$, $q$, $p - q'$ and $q'$ share the same properties,  there are essentially three types of integrals with $p^2$, $(p-q)^2$ and $(q'-q)^2$ in the numerator, which may be called as $K_1^{(1)}$, $K_1^{(2)}$ and  $K_1^{(3)}$, respectively. 

In appendix \ref{elements}, we examine these integrals.
We can show that there is no $1/\Delta$-pole term in the integrals of the type $K_1^{(1)}$, so we just neglect them.  For $K_1^{(2)}$ we examine an example of integral as
\begin{align}
K _{k,l,m,n} \frac{(\hat p)^{k+l+m+n}}{(p^2)^{2 - d/2 + \Delta}} \equiv  \int d^d q d^d q' \frac{(\hat p - \hat q)^k  ( \hat q )^l   ( \hat q' )^m (\hat p - \hat q')^n }{q^2 {q'}^2 (p - q' )^2 ((q - q')^2)^{d/2 - 1 + \Delta}} \, .
\label{Kklmn}
\end{align}
The $1/\Delta$-pole term is obtained as in \eqref{K12}. There are similar integrals with $q^2$, $(p-q')^2$, and ${q'}^2$ in the numerator, and they can be expressed by $K_{k,l,m,n}$ as in \eqref{exchange}.
For $K_1^{(3)}$, we define
\begin{align}
B_{k,l,m,n} \frac{(\hat p)^{k+l+m+n}}{(p^2)^{2 - d/2 + \Delta}}\equiv \int d^d q d^d q' \frac{(\hat p - \hat q)^k  ( \hat q )^l   ( \hat q' )^m (\hat p - \hat q')^n }{(p-q)^2 q^2  {q'}^2 (p - q' )^2 ((q - q')^2)^{d/2 - 2 + \Delta}} \, ,
\label{Bklmn}
\end{align}
whose $1/\Delta$-pole term can be deduced from the scalar case as in \eqref{B12}.

In terms of these elements of integral,  the integral $J_{k,l}$ in \eqref{Jp} can be written as
\begin{align}
 J _{k,l}= &
K_{s-l,l,s-k,k} - B_{k,s-k,l,s-l} + K_{s-k,k,s-l,l}- K_{l,s-1-l,k+1,s-k}  
- K_{s-1-l ,l ,s-k ,k+1} \\
& + K_{k+1 , s-1-k,l+1 , s- 1-l} - B_{k+1,s-k-1 ,l+1 , s-l-1}
+ K_{l+1,s-1-l,k+1,s-1-k} \nonumber \\ &
 - K_{s-1-k ,k ,s-l ,l+1}- K_{k, s-1-k, l+1 , s-l} \, . \nonumber
\end{align}
Using \eqref{pint} and
\begin{align}
 D_0^s = 2^{2s-1} N C_\phi^2 \Gamma (s+1) (2 \delta - 1 + s)_s \frac{( \hat x_{12})^{2s}}{(x_{12}^2)^{2s+ d - 2}}
 = (2 \delta )^{-2} s (d - 3 + s)  \tilde D_0^s \, ,
 \label{relationD}
\end{align}
we have
\begin{align}
\int d^d p \frac{(\hat p)^{2s}e^{ip \cdot x_{12}}}{(p^2)^{2-d/2}} 
 = (-1)^s \frac{\pi^{3d/2} 2^{2d-1} (d + 2s - 3) \Gamma(d - 2 + s)}{\delta^2 \Gamma(2 - d/2) (\Gamma(\delta))^2 N \Gamma (s)} \tilde D_0^s \, .
\end{align}
Then the integral $I_1^{(2)}$ in \eqref{I12p} becomes
\begin{align}
I_1^{(2)} &= (-1)^s  \frac{ d   (d + 2s - 3) \Gamma(d - 2 + s)}{ 4\pi^{d}\delta^2 \Gamma(2 - d/2) \Gamma(\delta) \Gamma (s)} \gamma_\psi \tilde D_0^s
\sum_{k,l=0}^{s-1} \tilde a_k \tilde a_l J_{k,l} \, .
\end{align}
Performing the summation over $k,l$ we find
\begin{align}
\label{resI12f}
 (I_1^{(2)})^\text{res} = \frac{d(d-2)}{2(s+d/2 -2)(s+d/2 -1)} \gamma_\psi \tilde D_0^s \, .
\end{align}
Combining the result in \eqref{resI11f}, the reside of $I_1$ at $\Delta = 0$ is given as in \eqref{fermionresult}.

\subsection{Integral $I_2$}
\label{threept}

For the integral $I_2$ we use the following three point function, which is evaluated with  the Wick contraction \eqref{Wickf} as
\begin{align}
\langle J_s^\epsilon (x_1 ) \tilde{\mathcal{O}} (x_2) \tilde{\mathcal{O}} (x_3) \rangle _0
 &= -\tilde N C_\phi^3 \sum_{k=0}^{s-1} \tilde  a_k \text{tr} \left[
  \hat \partial^k_1 \slashed{\partial}_3 \frac{1}{(x_{31}^2)^\delta} \hat \gamma \hat \partial^{s-1-k}_1 \slashed{\partial}_1 \frac{1}{(x_{12}^2)^\delta} \slashed{\partial}_2 \frac{1}{(x_{23}^2)^\delta} \right]+ (x_2 \leftrightarrow x_3)  \, .
  \nonumber 
\end{align}
We can rewrite
\begin{align}
\langle J_s^\epsilon (x_1 ) \tilde{\mathcal{O}}  (x_2) \tilde{\mathcal{O}} (x_3) \rangle _0 = 
P_s T_s (x_i ) 
\label{3ptp}
\end{align}
with 
\begin{align}
 T_s (x_i ) = 2 \tilde N  C_\phi^3
   \sum_{k=0}^{s-1} \tilde a_k  g_{4} 
     \hat \partial^k_1 \partial_1^{\mu_1} \frac{1}{(x_{31}^2)^\delta} \epsilon^{\mu_2} \hat \partial^{s-1-k}_1 \partial_1^{\mu_3} \frac{1}{(x_{12}^2)^\delta} \partial_2^{\mu_4} \frac{1}{(x_{23}^2)^\delta}  \, .
     \label{Ts}
\end{align}
Here we have used \eqref{tildeak}, \eqref{gammaab} and \eqref{Projector}.
It is convenient to use the following expression as (see appendix \ref{app3pt})
\begin{align}
T_s (x_i)=
   \frac{ 2^4 N \delta^2 C_\phi^3 }{(x_{23}^2)^{d/2}} \sum_{k=0}^s  a_k \left[ \hat \partial^k_1 \frac{1}{(x_{13}^2)^{\delta}} \right] \left[ \hat \partial^{s-k}_1 \frac{1}{(x_{12}^2)^{\delta} }\right] 
\label{fixed3pt}
\end{align}
with $a_k$  in \eqref{currentdef} used for the scalar case.

The integral  $I_2$ in \eqref{integrals} is now
\begin{align}
 I_2 =& 2^7 N^2 \delta^4 C_{\tilde \sigma}^2 C_\phi^6 P_s \int d^d x_3 d^d x_4 d^d x_5 d^d x_6 \sum_{k,l=0}^{s} a_k a_l \left[ \hat \partial^k _1 \frac{1}{(x_{13}^2)^{\delta}}\right]  \left[ \hat \partial^{s-k} _1 \frac{1}{(x_{14}^2)^\delta}\right]  \\
 & \qquad \times  \left[ \hat \partial^l _2 \frac{1}{(x_{25}^2)^{\delta}}\right]  \left[ \hat \partial^{s-l} _2 \frac{1}{(x_{26}^2)^{\delta}}\right]  \frac{1}{(x_{34}^2)^{d/2}} \frac{1}{(x_{56}^2)^{d/2}}\frac{1}{(x_{35}^2)^{1 - \Delta}}  \frac{1}{(x_{46}^2)^{1 - \Delta}}    \, . \nonumber
\end{align} 
The integral might be evaluated by applying the formula \eqref{unique}. However, we would meet divergence from a factor $a(d/2) = \Gamma (0)/\Gamma(d/2)$, where $d/2$ comes from the exponent of $x_{34}^2$ or $x_{56}^2$. In order to regularize the divergence, we shift the exponent by a small amount.
We choose to shift as
\begin{align}
 \label{tildeCkl}
&\tilde C_{k,l} = 
\int d^d x_3  d^d x_4 d^d x_5 d^d x_6 \left[ \hat \partial^{s-k}_1 \frac{1}{(x_{41}^2)^\alpha} \right]
  \left[ \hat \partial^{k}_1 \frac{1}{(x_{13}^2)^\alpha} \right] \\
&\times  \left[ \hat \partial^{s-l}_2 \frac{1}{(x_{62}^2)^\alpha} \right]
    \left[ \hat \partial^{l}_2 \frac{1}{(x_{25}^2)^\alpha} \right] \frac{1}{(x_{34}^2)^{\alpha + 1 + \Delta}} \frac{1}{(x_{56}^2)^{\alpha + 1}} \frac{1}{(x_{35}^2)^{\beta -1 - \Delta}} \frac{1}{(x_{46}^2)^{\beta -1 - \Delta}} \, .\nonumber
\end{align}
Here $\alpha = d/2 - 1 + \eta /2$ and $\beta = 2 - \eta$ as in \cite{Vasiliev:1981dg}, where $\eta/2$ physically corresponds to the anomalous dimension of $\psi^i$.
Expanding $\eta$ in $1/N$ as $\eta = \eta_0 + \eta_1 + \cdots$, the first two terms are given by $\eta_0/2 = 0$ and $\eta_1/2 = \gamma_\psi$ in \eqref{anopsi}.
Since we are working on the leading order in $1/N$, we set $\eta = \eta_0 = 0$ at the end of the computation.
Fortunately we will see that there is no divergence due to this procedure.
We also shift the exponent of $x_{34}^2$ as $\alpha + 1 \to \alpha + 1 +\Delta$ as discussed in subsection \ref{I2}.

We need to perform the integration over $x_3$, $x_4$, $x_5$ and $x_6$.
Using \eqref{unique} the integral over $x_3$ can be evaluated as
\begin{align}
\label{I2x3}
\hat \partial^{k}_1  \int d^d x_3  &
   \frac{1}{(x_{13}^2)^\alpha (x_{34}^2)^{\alpha + 1+ \Delta} (x_{35}^2)^{\beta - 1 - \Delta}} 
= v(\alpha , \alpha + 1 + \Delta , \beta - 1  - \Delta) \\
 &\qquad \times \frac{1}{(x_{45}^2)^{d/2 - \alpha}}  
 \sum_{a=0}^k \binom{k}{a}  \left[ \hat \partial^a_1  \frac{1}{(x_{15}^2)^{d/2 - \alpha - 1 - \Delta}} \right] \left[ \hat \partial^{k-a}_1  \frac{1}{(x_{14}^2)^{d/2 - \beta + 1 +\Delta}  } \right] \, . \nonumber
\end{align}
Note that $v(\alpha , \alpha +1 , \beta - 1) = \pi^{d/2} a(\alpha ) a(\alpha + 1) a(\beta - 1)$ diverges at $\eta =0$ as mentioned above.
The propagator between $x_1$ and $x_4$ can be rewritten as
\begin{align}
\label{kdep}
 &\left[ \hat \partial^{s-k}_1 \frac{1}{(x_{14}^2)^\alpha} \right]
\left[ \hat \partial^{k-a}_1  \frac{1}{(x_{14}^2)^{d/2 - \beta + 1 +\Delta} } \right] \\
& \quad =\frac{(\alpha)_{s-k} (d/2-\beta+1+\Delta)_{k-a}}{(d/2+\alpha-\beta + 1 +\Delta)_{s-a}}
\hat \partial^{s-a}_1 \frac{1}{(x_{14}^2)^{d/2 + \alpha - \beta + 1  + \Delta}} \, . 
\nonumber
\end{align}
Integration over $x_4$ then gives
\begin{align}
&\hat \partial^{s-a}_1 \int d^d x_4  \frac{1}{(x_{14}^2)^{d/2 + \alpha - \beta + 1  + \Delta} (x_{45}^2)^{d/2 - \alpha} (x_{46}^2)^{\beta - 1  - \Delta}}  \\
&=  v(d/2 + \alpha - \beta + 1  + \Delta , d/2 - \alpha , \beta - 1 - \Delta)  
\nonumber \\ & \qquad \qquad \times 
\frac{1}{(x_{56}^2)^{\beta - \alpha - 1  -\Delta}}  \sum_{b=0}^{s-a} \binom{s-a}{b} \left[ \hat \partial^{b}_1 
\frac{1}{ (x_{16}^2 )^{\alpha}} \right] \left[ \hat \partial^{s-a-b}_1  \frac{1}{ (x_{15}^2)^{d/2 - \beta + 1 + \Delta} } \right] \nonumber \, .
\end{align}
We rewrite the propagator between $x_1$ and $x_5$  as
\begin{align}
&\left[ \hat \partial^a_1  \frac{1}{(x_{15}^2)^{d/2 - \alpha - 1 - \Delta}} \right]
\left[ \hat \partial^{s-a-b}_1  \frac{1}{ (x_{15}^2)^{d/2 - \beta + 1 + \Delta} } \right] \\
& \qquad = \frac{\Gamma(a +d/2 - \alpha - 1 - \Delta) \Gamma(s -a - b + d/2 - \beta + 1 + \Delta) \Gamma (\alpha)}{\Gamma(d/2-\alpha-1-\Delta) \Gamma (d/2 - \beta + 1 + \Delta) \Gamma(s-b+\alpha)} 
\hat \partial^{s-b}_1  \frac{1}{ (x_{15}^2)^{\alpha} }\, . \nonumber
\end{align}
For $a \neq 0$, the divergence at $\eta =0$ is canceled out by the factor $1/\Gamma(d/2 - \alpha - 1)$. We will show that there is no contribution to the $1/\Delta$-pole even for $a=0$, and hence we can safely set $\eta = 0$ from now on.

The integral over $x_5$ and $x_6$ is the same as $B_{kl}$ defined in \eqref{Bkl} and evaluated as in \eqref{Bklsum}. Noticing \eqref{relationD}
we find
\begin{align}
& \tilde C_{k,l} = \frac{1}{\Delta} \frac{(2s + d - 3) \Gamma(d - 2 + s ) \pi^d \Gamma(2 - d/2)}{2^4 N C_\phi^4 \delta^{2} \Gamma(s) (\Gamma (\delta))^2 \Gamma(d/2)}  \tilde D_0^s \tilde H_2(k,l)   + \mathcal{O} (\Delta^0) \, ,
\end{align}
where
\begin{align}
\tilde  H_2 (k,l) =& \sum_{a=1}^k \sum_{b=0}^{s-a} \binom{k}{a} \binom{s-a}{b}  \\ & \times
 \frac{\Gamma(s - k + \delta) \Gamma (k - a + d/2 - 1) \Gamma (a) \Gamma (s- a - b + d/2 - 1)}{\Gamma (s - a + d - 2) \Gamma (s - b + \delta)} H_1^{(2)} (b ,l) 
 \, . \nonumber
\end{align}
Here $H_1^{(2)}$ was defined in \eqref{H12}.

\begin{figure}
 \centering
 \includegraphics[keepaspectratio, scale=0.5]
      {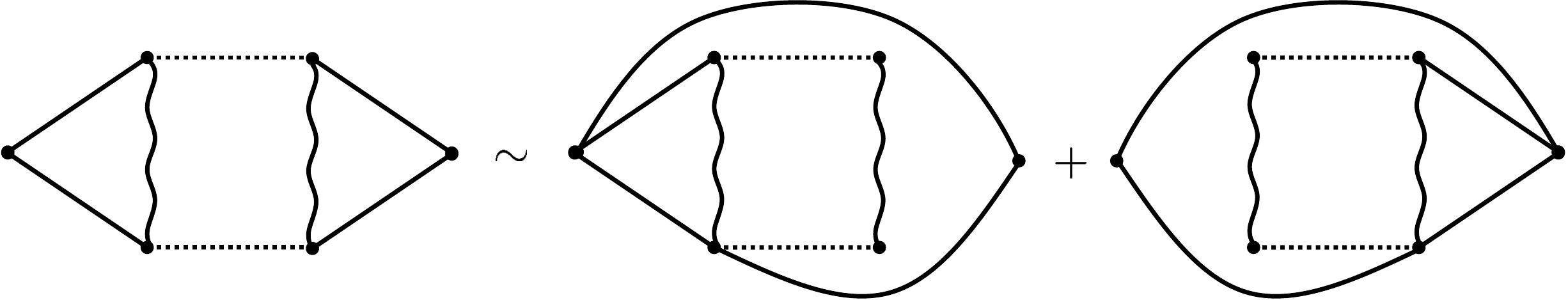}
 \caption{The left diagram represents the integral before introducing the shift of exponent by $\eta$ and $\Delta$.
 The solid line, the solid wavy line and the dotted line represent propagators with dimension $\delta$, $\delta + 1$ and $2 - \Delta$, respectively. Without the action of derivatives, we can show by closely following \cite{Vasiliev:1981dg} that the $1/\Delta$-pole term of the left diagram can be computed from the two right diagrams.}
 \label{3loop_0}
\end{figure}
As in the scalar case, the shift of exponent by $\Delta$ changes the overall factor of the $1/\Delta$-pole.
Without the action of derivatives, we can evaluate the residue before shifting the exponents as explained in fig.~\ref{3loop_0}. Comparing $\tilde C_{0,0}$ with $s=0$, we find that extra $1/2$ factor is needed.
The integral $I_2$ thus becomes
\begin{align}
\label{resI2f}
I_2 &=  2^7 N^2 \delta^4 C_{\tilde \sigma}^2 C_\phi^6 P_s \sum_{k,l =0}^s  a_k a_l \frac{1}{2} \tilde C_{k,l}  + \mathcal{O} (\Delta^0) \\
& = \frac{P_s}{\Delta} \frac{2 d \delta (2s + d - 3) \Gamma(d - 2 + s ) \Gamma(d-1) \gamma_\psi}{\Gamma(s) (\Gamma (\delta))^3  \Gamma(d/2)} \tilde D_0^s \sum_{k,l =0}^s a_k a_l  \tilde H_2(k,l)  + \mathcal{O} (\Delta^0) \, . \nonumber
\end{align}
After the summation over $k,l$ we obtain the expression in \eqref{fermionresult}.

Now it remains to show that there is no $1/\Delta$-pole for $a=0$ in \eqref{I2x3} at $\eta = 0$.
In the residue at $\Delta=0$ with $a=0$, $k$-dependence only appears in $\tilde a_k$ and some factors in \eqref{kdep}
as
\begin{align}
&\frac{1}{\eta} \Gamma \left(s-k+\frac{d}{2} -1 + \frac{\eta}{2} \right)\Gamma \left(k  + \frac{d}{2} - 1 + \eta \right)  
 \sim \frac{1}{\eta} \Gamma \left(s-k+\frac{d}{2} -1 \right) \Gamma \left(k  + \frac{d}{2} - 1 \right)  
 \nonumber \\  & \quad \times
\left[ 1 + \frac{\eta}{2} \psi \left(s - k + \frac{d}{2} -1 \right) + \eta  \psi \left(k  + \frac{d}{2} -1 \right) \right]  + \mathcal{O}(\eta^1) \, ,
\label{etafin}
\end{align}
where  $\psi (x) = \frac{d}{dx} \log \Gamma (x)$ as before.
Let us first study the term proportional to $1/\eta$.
For  $\Delta = 0$, the sum over $k$ reduces to
\begin{align}
&\sum_{k=0}^s a_k  \Gamma \left(s-k+\frac{d}{2} -1 \right)\Gamma \left(k  + \frac{d}{2} - 1 \right)   \\
& \quad  \propto \sum_{k=0}^s \frac{\Gamma (s+1)}{\Gamma (k+1) \Gamma (s-k+1)} (-1)^k = (1 - 1)^s = 0 \, ,
\nonumber
\end{align}
which implies that there is no term proportional to $1/\eta$ in the residue at $\Delta = 0$.
We then examine finite contributions at $\eta=0$.
They would come from the second and third terms in the second line of \eqref{etafin}.
However, we can show that the summation over the other variables $b,l$ leads to
\begin{align}
\label{resa0}
\sum_{l=0}^s  a_l \sum_{b=0}^s \binom{s}{b}  H_{1}^{(2)} (b,l) = 0 \, .
\end{align}
This confirms our claim that  there is no $1/\Delta$-pole for $a=0$ in \eqref{I2x3} at $\eta = 0$.

\section{Generalizations}
\label{General}

In the previous sections, we have studied higher spin currents in the singlet of global symmetry for the critical O$(N)$ scalar model and the Gross-Neveu model. There are several examples, where we can deduce the anomalous dimensions without extra efforts.
In the next subsection, we apply the analysis to higher spin currents in the non-singlet representations with respect to O$(N)$ or U$(\tilde N)$ global symmetry.
In subsection \ref{UNscalar}, we compute the anomalous dimensions in the system of U$(\tilde N)$ complex scalars.
In subsection \ref{super}, we examine  the $\mathcal{N}=2$ supersymmetric U$(\tilde N)$ model, and in particular, we examine the coincidence of anomalous dimensions among 3d theories of  bosons and fermions.

\subsection{Non-singlet currents}
\label{non-singlet}

In the free scalar theory, there are many conserved currents as
\begin{align}
J_s^{\epsilon , ij} = \sum_{k=0}^s a_k \hat \partial^{s-k} \phi^i \hat \partial^k \phi^j
\end{align}
with $a_k$ in \eqref{currentdef}. According to the O$(N)$ representation, we divide them into three types as $J^{\epsilon,(ij)}_s$ for the symmetric traceless one and $J^{\epsilon,[ij]}_s$ for  the antisymmetric one along with $J^\epsilon_s $ in \eqref{currentdef} for the singlet one. 
Up to now we have focused on the single currents since they are dual to  higher spin gauge fields.
However, our methods are not restricted to the singlet ones but can be applied more generically. In fact, the anomalous dimensions for the other types of currents can be deduced from the analysis for the singlet as seen below.

The two point function of the currents are
\begin{align}
&\langle J_s^{\epsilon,(ij)} (x_1) J_s^{\epsilon,(kl)} (x_2) \rangle_0 =  (\delta^{il} \delta^{jk} + \delta^{ik} \delta^{jl} - \frac{2}{N} \delta^{ij} \delta^{kl}) \frac{D_0^s}{2N} \, , \\
&\langle J_s^{\epsilon,[ij]} (x_1) J_s^{\epsilon,[kl]} (x_2) \rangle_0 =  (\delta^{il} \delta^{jk} - \delta^{ik} \delta^{jl} ) \frac{D_0^s}{2N}
\end{align}
with $D_0^s$ defined in \eqref{D0s}.
We would like to compute the anomalous dimensions of these currents at the critical point.
For the purpose we need the information of three and four point functions involving these currents.
We can easily see that three point functions with non-singlet current vanish, so there are no contributions of the type $I_2$. The four point functions are related to that with singlet current in \eqref{fourpt} as
\begin{align}
&\langle J_s^{(ij)} (x_1) J_s^{(kl)} (x_2) \mathcal{O}(x_3) \mathcal{O}(x_4) \rangle_0 =  (\delta^{il} \delta^{jk} + \delta^{ik} \delta^{jl} - \frac{2}{N} \delta^{ij} \delta^{kl}) \frac{K_1 (x_i) + K_2 (x_i)}{2N} \, , \\
&\langle J_s^{[ij]} (x_1) J_s^{[kl]} (x_2) \mathcal{O}(x_3) \mathcal{O}(x_4) \rangle_0 =   (\delta^{il} \delta^{jk} - \delta^{ik} \delta^{jl} ) \frac{K_1 (x_i) + K_2 (x_i)}{2N} \, .
\end{align}
Therefore, the anomalous dimensions are
\begin{align}
\gamma_{s(ij)} = \gamma_{s[ij]} = 2 \gamma_\phi \frac{(s-1)(s + d - 2) }{ (s + d/2 - 2) (s + d/2 - 1)}  \, ,
\end{align}
which reproduces the known results \cite{Lang:1992zw}.

As in the free scalar theory, there are also non-singlet currents in the free fermion theory as
\begin{align}
J_s^{\epsilon,ij} (x ) = \sum_{k=0}^{s-1} \tilde a_k \hat \partial^k \bar \psi ^i  \hat \gamma \hat \partial ^{s-k-1} \psi ^ j 
\end{align}
with $\tilde a_k$ in \eqref{tildeak}.
The two point functions are
\begin{align}
&\langle J_s^{(ij)} (x_1) J_s^{(kl)} (x_2) \rangle_0 =  (\delta^{il} \delta^{jk} + \delta^{ik} \delta^{jl} - \frac{2}{\tilde N} \delta^{ij} \delta^{kl}) \frac{\tilde D_0^s}{\tilde N} \, , \\
&\langle J_s^{[ij]} (x_1) J_s^{[kl]} (x_2) \rangle_0 =  (\delta^{il} \delta^{jk} - \delta^{ik} \delta^{jl} ) \frac{\tilde D_0^s}{\tilde N}
\end{align}
with $\tilde D_0^s$ in \eqref{2pt}.
The three point functions with a non-singlet current vanishes, and the four point functions are
\begin{align}
&\langle J_s^{(ij)} (x_1) J_s^{(kl)} (x_2) \mathcal{O}(x_3) \mathcal{O}(x_4) \rangle_0 =  (\delta^{il} \delta^{jk} + \delta^{ik} \delta^{jl} - \frac{2}{\tilde N} \delta^{ij} \delta^{kl}) \frac{\tilde K_1 (x_i) + \tilde K_2 (x_i)}{\tilde N} \, , \\
&\langle J_s^{[ij]} (x_1) J_s^{[kl]} (x_2) \mathcal{O}(x_3) \mathcal{O}(x_4) \rangle_0 =   (\delta^{il} \delta^{jk} - \delta^{ik} \delta^{jl} ) \frac{\tilde K_1 (x_i) + \tilde K_2 (x_i)}{\tilde N} \, .
\end{align}
Here $\tilde K_a (x_i)$ are defined in \eqref{4ptp1} and \eqref{4ptp2}.
Therefore, the anomalous dimensions are
\begin{align}
\gamma_{s(ij)} = \gamma_{s[ij]} = 2 \gamma_\psi \frac{(s-1)(s + d - 2) }{ (s + d/2 - 2) (s + d/2 - 1)} 
\end{align}
as  found in \cite{Muta:1976js}.

\subsection{U$(\tilde N)$ scalar model}
\label{UNscalar}

In this paper, we have considered the system of O$(N)$ real scalars, but it is not difficult to generalize to that of U$(\tilde N)$ complex scalars. The results for  U$(\tilde N)$ scalars will be utilized for the arguments on the $\mathcal{N}=2$ supersymmetric setup in the next subsection.

The U$(\tilde N)$ scalar model consists of $\tilde N$ complex scalars, and we use the following normalization as 
\begin{align}
\langle {\phi}^i (x_1) {\phi^*}^j (x_2) \rangle= C_\phi 
\frac{\delta^{ij}}{(x_{12}^2)^{\delta}}\label{Wickc}
\end{align}
with $C_\phi$ in \eqref{Wicks}. The system has conserved currents, which can be given by
\begin{align}
 J_s ^\epsilon = \sum_{k=0}^s a_k \hat \partial^{s-k} {\phi^*}^i \hat \partial^k \phi_i 
\, , \quad \langle J_s^{\epsilon} (x_1) J_s^{\epsilon} (x_2) \rangle =  \frac{\tilde N}{2N} D_0^s \, .
\end{align}
We deform the system by \eqref{def} with the scalar operator
\begin{align}
{\mathcal{O}}_{2 \delta} (x) \equiv
\mathcal{O} (x) = {\phi^*}^i \phi_i \, , \quad 
\langle \mathcal{O}(x_1) \mathcal{O} (x_2) \rangle = 
 \tilde N C_\phi^2 \frac{1}{(x_{12}^2)^{2 \delta} }\, .
 \label{UNso}
\end{align} 
The propagator \eqref{Greg} is given by \eqref{sigmapro}, but $C_{\mathcal{O}}$ is with extra factor 2 and with $N$ replaced by $\tilde N$.

We would like to examine how the anomalous dimensions of higher spin currents change from the O$(N)$ scalar case. We can see that the four and three point functions in \eqref{fourpt} and \eqref{threepts} become $1/8$ and $1/4$ times along with $N$ replaced by $\tilde N$. 
Combining with the changes in the current-current two point function and the propagator \eqref{Greg}, the anomalous dimensions are given by a half of \eqref{anomalouss} with $N$ replaced by $\tilde N$. Denoting the anomalous dimension of spin $s$ current as $\gamma_s^{\text{U}(\tilde N)}$ and that for the Gross-Neveu model in \eqref{anomalousf} as $\gamma_s^\text{GN}$, we have the relation as
\begin{align}
 \gamma_s^{\text{U}(\tilde N)} = \frac{\text{tr}\mathbbm{1} (4-d)}{2 (d-2)}  \gamma_s^\text{GN} \, .
 \label{coincidence}
\end{align}
For $d=3$, $\text{tr} \mathbbm{1} = 2$ and thus we find $\gamma_s^{\text{U}(\tilde N)} = \gamma_s^\text{GN}$.

\subsection{$\mathcal{N}=2$ supersymmetric model in 3 dimensions}
\label{super}

Comparing the results from the O$(N)$ (U$(\tilde N)$) scalar model in \eqref{anomalouss} and from the Gross-Neveu model in \eqref{anomalousf}, we can see that they are almost identical except for the overall $d$-dependent factor.
In particular, we observe the coincidence for $d=3$ as in \eqref{coincidence}.
We do not have a clear explanation of this fact in general $d$, but we may be able to explain it as a consequence of $\mathcal{N}=2$ supersymmetry for $d=3$.

For $d=3$, a free theory with $\mathcal{N}=2$ supersymmetry can be constructed from the sum of $\tilde N$ complex bosons and $\tilde N$ Dirac fermions. We may consider the following marginal deformation as
\begin{align}
\lambda \int d^d x \mathcal{O} (x) \tilde{\mathcal{O}}(x) 
\label{marginal}
\end{align}
plus supersymmetric completion with the product of fermionic operators.
Here $\mathcal{O} (x)$ and $\tilde{\mathcal{O}}(x)$ are the scalar operators in \eqref{UNso} and \eqref{GNso}, respectively.
The fixed line under this deformation can be understood  from the dual gravity theory as argued in \cite{Leigh:2003gk} at least for large $N$ as explained below.

The scalar operators $\mathcal{O} (x)$ and $\tilde{\mathcal{O}}(x)$ are dual to complex scalar fields $\varphi$ and $\varphi '$ with Neumann and Dirichlet boundary conditions. The marginal deformation in \eqref{marginal} mixes the boundary conditions of these two complex scalars 
\cite{Witten:2001ua}, see also \cite{Chang:2012kt,Hikida:2015nfa}. In order to preserve the supersymmetry, we also need to change the boundary conditions for two Dirac fermions as well, and this was analyzed in the current context in details in \cite{Chang:2012kt}. At the limit of $\lambda \to \infty$, the boundary conditions are exchanged such that Dirichlet and Neumann ones are assigned for $\varphi$ and $\varphi '$.
This bulk theory is dual to the sum of critical U$(\tilde N)$ scalar model and the critical Gross-Neveu model plus interactions mixing the two models, and it is natural to expect that the system has $\mathcal{N}=2$ supersymmetry at least for large $N$ as discussed in \cite{Leigh:2003gk} for $\mathcal{N}=1$ setup.
Performing a projection, we would obtain the bulk theory dual to the  critical U$(\tilde N)$ scalar model or the critical Gross-Neveu model.
The original $\mathcal{N}=2$ supersymmetry relates the two models, and this may 
explain the coincidence of anomalous dimensions as observed in \eqref{coincidence} with $d=3$.
We hope to be able to report on more details in near future.

\section{Conclusion and discussions}
\label{conclusion}

In this paper, we have examined the critical O$(N)$ scalar model and the Gross-Neveu model in arbitrary $d$ dimensions, which are proposed to be dual to the type A and type B Vasiliev theories, respectively.
We have first extended the analysis on 3d O$(N)$ scalar in \cite{Hikida:2016wqj} to the case with generic dimension $d$. 
In particular, we use a different regularization by shifting the exponent of propagator as in \eqref{Greg}. Moreover, we have developed a new way to evaluate the effects of derivatives acting on the propagators. Using these methods, we have reproduced the known results on anomalous dimensions of higher spin currents to the $1/N$ order in \cite{Lang:1992zw}, see \eqref{anomalouss}. 
We then apply these methods to the Gross-Neveu model in $d$ dimensions.
In this case, the introduction of $\Delta$  in \eqref{Greg} is not enough to regularize the divergence in a Feynman integral. We shift the dimension of fermion field by $\eta/2$, and then take the limit $\eta \to 0$ at the end of the computation. The limit is found to be finite, and we can reproduce the previous result on anomalous dimensions to the $1/N$ order in \cite{Muta:1976js}. They are the same as those for the scalar case up to a $d$-dependent factor, see \eqref{anomalousf}.
It is an interesting question why the coincidence happens, which may be answered by using higher spin symmetry. For $d=3$, we give a possible explanation of it as a consequence of $\mathcal{N}=2$ supersymmetry.

We investigated these two models to understand the dual bulk mechanism of symmetry breaking.
The anomalous dimensions can be computed by making use of bulk Witten diagrams, and the diagrams can be explained in terms of boundary conformal perturbation theory as in \cite{Hikida:2016wqj}, see also \cite{Hartman:2006dy,Giombi:2011ya}.
Therefore, the conclusion obtained in \cite{Hikida:2016wqj} can be applied also to the current examples. Namely, the boundary computation confirms the bulk interpretations such that the mass arises from a one-loop effect due to the change of boundary condition for the bulk scalar field. For the type A Vasiliev theory on AdS$_4$, the Goldstone modes are identified with bound states of scalar and higher spin field through group theoretic argument in \cite{Girardello:2002pp}. However, the boundary computation in  \cite{Hikida:2016wqj} could not explain what are the Goldstone modes. The analysis in this paper does not add anything new on this issue, so we again cannot identify the Goldstone modes from the boundary computations. In order to confirm quantitatively the argument in \cite{Girardello:2002pp} or its generalizations, we may have to directly evaluate the bulk loop diagrams.

There are other open problems as follows;
The system of free fermions in $d >3$ dimensions has mixed symmetric conserved currents,
and it is interesting to extend the current analysis to these currents.
Moreover, as in subsection \ref{super}, the analysis for the Gross-Neveu model should be applicable to $\mathcal{N}=1$ and $\mathcal{N}=2$ supersymmetric cases as in \cite{Leigh:2003gk,Sezgin:2003pt}.
In order to examine the relation to superstring theory,
it important to couple the bosonic and fermionic fields to Chern-Simons gauge fields  \cite{Aharony:2011jz,Giombi:2011kc,Maldacena:2012sf} as explained in the introduction.
We think that the conformal perturbation theory is suitable for the extensions mentioned above,
but it should be useful to examine the relations to other methods particularly for O$(N)$ scalars, e.g., in \cite{Lang:1992zw,Skvortsov:2015pea,Giombi:2016hkj,Li:2015rfa,Alday:2015ewa,Dey:2016zbg}.%
\footnote{It might be also interesting to examine O$(N)$ symmetry breaking, see \cite{Leigh:2012mz,Alvarez-Gaume:2016vff} for examples.}
As a technical issue, we utilized several sum formulas to obtain the results on the anomalous dimensions in \eqref{anomalouss} and \eqref{anomalousf}, but the formulas are checked by Mathematica for several explicit examples. It is desired to obtain analytic proofs, see also footnote \ref{Mathematica}.

\subsection*{Acknowledgements}

The work of YH is supported in part by JSPS KAKENHI Grant Number 24740170, 16H02182.

\appendix

\section{Alternative method for $B_{k,l}$}
\label{alt}

In subsection \ref{I12}, $B_{k,l}$ in \eqref{Bkl} has been computed as \eqref{Bklsum} with \eqref{H12} in the momentum representation. In this appendix, we compute $B_{k,l}$ with the method developed in section \ref{I2}. Instead of $B_{k,l}$ in \eqref{Bkl}, we evaluate
 \begin{align}
 B_{k,l} ' =  \int d^d x_3 d^d x_4  \left[ \hat \partial^k_1  \frac{1}{(x_{13}^2)^{\delta + \Delta/2} } \right] \left[ \hat \partial^{s - l}_2 \frac{1}{(x_{32}^2)^{\delta + \Delta/2} } \right] \left[ \hat \partial^{s-k}_1  \frac{1}{(x_{41}^2)^{\delta} } \right]
 \left[ \hat \partial_2 ^{l} \frac{1}{(x_{24}^2)^{\delta }} \right]  \frac{1}{(x_{34}^2)^{2 - \Delta}} \label{tBkl}
\end{align}
with the shifts of exponent in the propagators. The effect of derivatives is the same as $B_{k,l}$ unless extra zeros or poles appear at $\Delta = 0$. However, the overall factor would changes, which will be fixed by comparing the result in \eqref{Bklsum} for $s=0$.

First we perform the integral over $x_3$ as
 \begin{align}
&\hat \partial^{k}_1 \hat \partial_2 ^{s-l} \int  d^d x_3   \frac{1}{(x_{31}^2)^{\delta + \Delta/2} (x_{23}^2)^{\delta + \Delta / 2 } (x_{34}^2)^{2 - \Delta}}  \\
&= v(\delta + \Delta/2 , \delta + \Delta/2 , 2 - \Delta ) \hat \partial^{k}_1 \hat \partial_2 ^{s-l}
\frac{1}{(x_{24}^2)^{1 - \Delta /2} (x_{14}^2)^{1 - \frac{\Delta}{2}} (x_{12}^2)^{d/2 - 2 + \Delta}} \nonumber \\
&= \pi^{d/2} \frac{(\Gamma(1 - \Delta / 2))^2 \Gamma(d/2 - 2 + \Delta)}{(\Gamma (\delta + \Delta / 2 ))^2 \Gamma (2 - \Delta)} \nonumber \\ 
& \qquad \times
\sum_{a=0}^{k} \sum_{b=0}^{s-l} \binom{k}{a} \binom{s-l}{b}
\left[\hat \partial_2 ^{s-l-b} \frac{1}{(x_{24}^2)^{1 - \Delta /2}} \right] \left[  \hat \partial^{k-a}_1 \frac{1}{ (x_{14}^2)^{1 - \Delta / 2} }  \right] \left[ \hat \partial^{a}_1  \hat \partial_2 ^{b} \frac{1}{ (x_{12}^2)^{d/2 - 2 + \Delta}} \right] \nonumber \, .
\end{align}
Using
\begin{align}
&\left[ \hat \partial^{s-k}_1  \frac{1}{(x_{14}^2)^{\delta} } \right] \left[  \hat \partial^{k-a}_1 \frac{1}{ (x_{14}^2)^{1 - \Delta / 2} }  \right]  \nonumber \\ & \quad
= \frac{\Gamma (s-k+\delta) \Gamma(k-a+ 1 - \Delta / 2) \Gamma(d / 2 - \Delta / 2 )}{\Gamma(\delta) \Gamma(1 - \Delta / 2) \Gamma(s - a + d / 2 - \Delta / 2 )}
\hat \partial^{s-a}_1  \frac{1}{(x_{14}^2)^{d/2- \Delta /2} } \, , \\
& \left[ \hat \partial_2 ^{l} \frac{1}{(x_{24}^2)^{\delta }} \right] \left[\hat \partial_2 ^{s-l-b} \frac{1}{(x_{24}^2)^{1 - \Delta /2}} \right] \nonumber
\\ & \quad =  \frac{\Gamma (l+\delta) \Gamma(s-l-b+ 1 - \Delta / 2 ) \Gamma(d/2 - \Delta / 2)}{\Gamma(\delta) \Gamma(1 - \Delta / 2) \Gamma(s - b + d / 2 - \Delta / 2 )}
\hat \partial_2 ^{s-b} \frac{1}{(x_{24}^2)^{d/2 - \Delta /2}} \nonumber \, ,
\end{align}
the integral over $x_4$ becomes
\begin{align}
&\hat \partial^{s-a}_1 \hat \partial_2 ^{s-b}  \int d^d x_4 \frac{1}{(x_{14}^2)^{d/2- \Delta /2} (x_{24}^2)^{d/2 - \Delta /2}}  
= v (d / 2 - \Delta / 2 ,d / 2 - \Delta / 2  , \Delta )\hat \partial^{s-a}_1 \hat \partial_2 ^{s-b}  \frac{1}{(x_{12}^2)^{d/2 - \Delta}} \nonumber \\
&= \frac{1}{\Delta} \frac{4 \pi^{d/2}}{\Gamma(d/2)}\hat \partial^{s-a}_1 \hat \partial_2 ^{s-b} \frac{1}{(x_{12}^2)^{d/2 - \Delta}} + \mathcal{O}(\Delta^0) \, .
\end{align}
Rewriting as
\begin{align}
&\left[ \hat \partial^{a}_1  \hat \partial_2 ^{b} \frac{1}{ (x_{12}^2)^{d/2 - 2 + \Delta}} \right]
\left[\hat \partial^{s-a}_1 \hat \partial_2 ^{s-b} \frac{1}{(x_{12}^2)^{d/2 - \Delta}} \right]  \\
& = (-1)^s 2^{2s} \frac{\Gamma(a+b + d/2 - 2 + \Delta) \Gamma (2s -a - b + d/2- \Delta)}{\Gamma(d/2 - 2 + \Delta) \Gamma (d/2- \Delta)} 
\frac{(\hat x_{12})^{2s}}{(x_{12}^2)^{2 \delta + 2s}} \nonumber \\
& = (-1)^s  \frac{\Gamma(a+b + d/2 - 2 + \Delta) \Gamma (2s -a - b + d/2 - \Delta)}{\Gamma(d/2 - 2 + \Delta) \Gamma (d/2 - \Delta)} 
 \frac{2}{C_\phi^2 N \Gamma (s+1) (2 \delta - 1 + s)_s} D_0^s \, , \nonumber
\end{align}
we find
\begin{align}
 B_{k,l} ' = \frac{1}{\Delta} \frac{\Gamma (2 \delta - 1 + s)}{2 C_\phi^4 N \Gamma(s+1) (\Gamma (\delta))^2 \Gamma(2 \delta - 1 + 2s)} D_0^s  {H_1^{(2)}}' (k,l)  + \mathcal{O} (\Delta^0)
\end{align}
with
\begin{align}
 {H_1^{(2)}}' (k,l)  &= (-1)^s \sum_{a=0}^k \sum_{b=0}^{s-l} \binom{k}{a} \binom{s-l}{b} 
 \frac{\Gamma (s - k + \delta) \Gamma(k -a + 1) \Gamma ( l + \delta) \Gamma (s-l - b + 1) }{\Gamma (s - a + d/2) \Gamma (s - b + d/2)} \nonumber \\ & \qquad \times
 \Gamma (a + b + d/2 - 2) \Gamma (2s - a - b + d/2) \, . 
\end{align}
Here $D_0^s$ is defined as \eqref{D0s} with \eqref{current2pt} and \eqref{current2ptCs}.
Comparing $B_{0,0}$ with $s=0$ in \eqref{Bklsum}, we find
\begin{align}
 B_{00} = \frac{1}{2} B_{00} ' + \mathcal{O}(\Delta^0) \, . 
\end{align}
In this way we arrive at an alternative expression of $B_{k,l}$ as
\begin{align}
 B_{k.l} = \frac{1}{\Delta} \frac{\Gamma (2 \delta - 1 + s)}{4 C_\phi^4 N \Gamma(s+1) (\Gamma (\delta))^2 \Gamma(2 \delta - 1 + 2s)}D_0^s {H_1^{(2)}}' (k,l)  + \mathcal{O} (\Delta^0) \, .
\end{align}
This implies that 
\begin{align}
I_1^{(2)} = \frac{1}{\Delta} \frac{2 \Gamma (2 \delta - 1 + s)}{ \Gamma(s+1) (\Gamma (\delta))^2 \Gamma(2 \delta - 1 + 2s)} C_\sigma D_0^s \sum_{k,l=0}^s a_k a_l { H_1^{(2)} } ' (k,l)  + \mathcal{O} (\Delta^0) \, ,
\label{altI12}
\end{align}
and we can check that the sum over $k,l$ reproduces \eqref{resI12} for explicit values of $d,s$.

\section{Technical details for the theory of fermions}
\label{technical}

In this appendix we collect some details of computation in the fermion theory.

\subsection{Two point function}
\label{app2pt}
We would like to check  the relation \eqref{formula1} by following the analysis in appendix B of \cite{Diaz:2006nm}.
The summation over $k$ can be rewritten as
\begin{align}
 \sum_{k = 0}^{s-1} 
a_k (\delta)_{s-k+l} (\delta)_{s+k-l} &=
\sum_{k=0}^{s-1} \frac{(1-s)_k}{k!} \frac{(1-s-\delta)_k (s-l+\delta)_k}{(\delta + 1)_k (1 - \delta - s - l)_k} (\delta )_{s-l}(\delta )_{s+l} \\
& =
{}_3 F_2 
\binom{1-s, 1-s-\delta ,s-l+\delta}{\delta+1, 1-\delta -s-l} (\delta )_{s-l}(\delta )_{s+l} \, . \nonumber
\end{align}
 Using the formulas
\begin{align}
& _3F_2 
\binom{-n, a,b}{d,e}
=\frac{(e-a)_n}{(e)_n}
\ _3F_2 
\binom{-n, a, d-b}{d, 1-n+a-e} \nonumber 
\end{align}
and
\begin{align}
& {}_3F_2 
\binom{-n, a,b}{c,a}
=
\ {}_2F_1 
\binom{-n, b}{c} =\frac{(c-b)_n}{(c)_n} \, , \quad  (n \in \mathbb N) \, ,  \label{formula2F1}
\end{align}
we obtain
\begin{align}
 {}_3F_2 
\binom{1-s, 1-s-\delta ,s-l+\delta}{\delta+1, 1-\delta -s-l}
&
=
\frac{(1-s+l)_{s-1}}{(\delta +1)_{s-1}}
\frac{(1-s-\delta )_{l}}{(1-l)_{l}}
{}_3F_2 
\binom{ -l, s-l+\delta, - \delta -l}{1-\delta -s-l, s-l+\delta} 
 \nonumber
\\
&
=
\frac{(1-s+l)_{s-1}}{(\delta +1)_{s-1}}
\frac{(1-s-\delta )_{l}}{(1-l)_{l}}
\frac{(1-s)_{l}}{(1-\delta -s-l)_{l}} \,  .
\label{k-sum}
\end{align}
As a result, left hand side of \eqref{formula1} becomes
\begin{align} 
\sum_{k,l = 0}^{s-1} (-1)^{s+1}
     a_k a_l (\delta)_{s-k+l} (\delta)_{s+k-l} 
&=
\sum_{l = 0}^{s-1} 
\frac{(1-s)_l}{l!}
\frac{(1-s-\delta )_l}{(\delta +1)_l}
\frac{\Gamma (\delta +1 )}{(s)_\delta}
(\delta)_{s} (\delta)_{s} \\
&= 
{}_2F_1\binom{1-s, 1-s-\delta }{\delta +1}
\frac{\Gamma (\delta +1 )}{(s)_\delta}
(\delta)_{s} (\delta)_{s} \, .\nonumber 
\end{align}  
Applying the formula \eqref{formula2F1}, we find the relation \eqref{formula1}.

\subsection{Three point function}
\label{app3pt}

In this appendix we derive \eqref{fixed3pt}.
Noticing $\slashed{\epsilon} \slashed{\epsilon}  = 0 $, $T_s (x_i)$ in \eqref{Ts} can be rewritten as
\begin{align}
T_s (x_i)
  = 2 \tilde N (d-2)^3 C_\phi^3 \text{tr} (\slashed{x}_{31} \slashed{\epsilon} \slashed{x}_{12}  \slashed{x}_{23})
    \frac{1}{(x_{23}^2)^{d/2}}
   \sum_{k=0}^{s-1} \tilde a_k
    \left[ \hat \partial_1^k \frac{1}{(x_{31}^2)^{d/2}} \right]   
    \left[ \hat \partial_1^{s-1-k} \frac{1}{(x_{12}^2)^{d/2}} \right]  \, . 
\end{align}
Using \eqref{gammatrace}, we find
\begin{align}
& \frac{1}{\text{tr} \mathbbm{1}} \text{tr} (\slashed{\epsilon} \slashed{x}_{12} \slashed{x}_{23} \slashed{x}_{31}) 
= \hat x_{12} x_{23} \cdot x_{31} - \hat x_{23} x_{12} \cdot x_{31} + \hat x_{31} x_{12} \cdot x_{23} \\
&=  \tfrac12 \left[ \hat x_{12} (x_{12}^2 - x_{23}^2 -  x_{31}^2 )- \hat x_{23} ( x_{23}^2 - x_{12}^2 -  x_{31}^2 )  + \hat x_{31} ( x_{13}^2 -  x_{12}^2 -  x_{23} ^2 ) \right] \nonumber \\
&= x_{12}^2 \hat x_{13} -  x_{13}^2 \hat x_{12} 
=x_{12}^2 x_{13}^2 \left( \frac{\hat x_{13}}{x_{13}^2} - \frac{\hat x_{12}}{x_{12}^2 }  \right) \, .  \nonumber 
\end{align}
Moreover, we have
\begin{align}
\label{3ptform0}
 &  \sum_{k=0}^{s-1} \tilde  a_k
    \left[ \hat \partial_1^k \frac{1}{(x_{31}^2)^{d/2}} \right]   
    \left[ \hat \partial_1^{s-1-k} \frac{1}{(x_{12}^2)^{d/2}} \right]  \\
 &   = \sum_{k=0}^{s-1} (-1)^k \frac{\Gamma (s) \Gamma(s + d/2 - 1)}{\Gamma(k+1) \Gamma(s -k) \Gamma (d/2)}
  \frac{(2 \hat x_{31})^k }{(x_{31}^2)^{d/2 + k}} \frac{(2 \hat x_{21})^{s-1-k} }{(x_{21}^2)^{s-1 -k+d/2}} \nonumber  \\
& = \frac{2^{s-1} \Gamma(s + d/2 - 1)}{ \Gamma (d/2)} \left( \frac{\hat x_{13}}{x_{13}^2} - \frac{\hat x_{12}}{x_{12}^2 }  \right)^{s-1} \frac{1}{(x_{13}^2)^{d/2}}  \frac{1}{(x_{12}^2)^{d/2}}  \nonumber
\end{align}
and
\begin{align}
\label{3ptform1}
 &  \sum_{k=0}^{s} a_k
    \left[ \hat \partial_1^k \frac{1}{(x_{31}^2)^\delta } \right]   
    \left[ \hat \partial_1^{s-k} \frac{1}{(x_{12}^2)^\delta } \right]  \\
 &   = \sum_{k=0}^{s-1}\frac{ (-1)^k }{2} \frac{\Gamma (s+1) \Gamma(s + d/2 - 1)}{\Gamma(k+1) \Gamma(s -k+1) \Gamma (\delta)}
  \frac{(2 \hat x_{31})^k }{(x_{31}^2)^{\delta + k}} \frac{(2 \hat x_{21})^{s-k} }{(x_{21}^2)^{\delta + s -k}} \nonumber  \\
& = \frac{2^{s-1} \Gamma(s + d/2 - 1)}{ \Gamma (\delta)} \left( \frac{\hat x_{13}}{x_{13}^2} - \frac{\hat x_{12}}{x_{12}^2 }  \right)^{s} \frac{1}{(x_{13}^2)^{\delta}}  \frac{1}{(x_{12}^2)^{\delta}}  \, .\nonumber
\end{align}
Here $a_k$ was introduced in \eqref{currentdef} for the O$(N)$ scalars.
Using these equations we obtain \eqref{fixed3pt}.

\subsection{Elements for integral $I_1^{(2)}$}
\label{elements}

In subsection \ref{I12f}, we evaluate the integral $I_1^{(2)}$ in the theory of fermions.
For the purpose we have divided the integral into several pieces, which are classified into three types  as  $K_1^{(a)}$ with $a=1,2,3$.
We are interested in singular terms at $\Delta = 0$.
We shall show that there is no such a term for the integral of the type $K_1^{(1)}$.
For the type $K_1^{(3)}$  we may move back from the momentum representation to the coordinate one.
Then we can show the integral is the same type as that for the O$(N)$ scalars.
Therefore, the new non-trivial integrals are of the type $K_1^{(2)}$.

We start to show that there is no $1/\Delta$-pole term for $K_1^{(1)}$.
We examine
\begin{align}
 \int d^d q d^d q' \frac{1}{(p-q)^2 q^2 {q'}^2 (p - q' )^2 ((q - q')^2)^{d/2 - 1 +  \Delta} }
\end{align}
for simplicity. It is straightforward to show that the same is true also for the cases with $\hat q$ and so on in the numerator.
The integral involving $q$ can be rewritten as
\begin{align}
\label{K11q}
&\int d^d q \frac{1}{(p-q)^2 q^2 ((q - q')^2)^{d/2 - 1 +  \Delta}}
  = \frac{\Gamma (d/2 + 1 + \Delta )}{\Gamma (\delta + \Delta)}\int d^d q
\int_0^1 dy \int_0^{1-y} dx \\
& \qquad \qquad \times \frac{y^{d/2 - 2 + \Delta}}{((q - xp -yq')^2 + x(1-x)p^2 + y(1-y){q'}^2 - 2 xy p \cdot q')^{d/2 + 1 + \Delta}} \, . \nonumber 
\end{align}
Integration over $q$ then leads to
\begin{align}
& \pi^{d/2} \frac{\Gamma (1 + \Delta )}{\Gamma (\delta + \Delta)}
\int_0^1 dy \int_0^{1-y} dx \frac{y^{d/2 - 2 + \Delta}}{( x(1-x)p^2 + y(1-y){q'}^2 - 2 xy p \cdot q')^{1 + \Delta}} \nonumber \\
&= \pi^{d/2} \frac{\Gamma (1 + \Delta )}{\Gamma (\delta + \Delta)}
 \int_0^1 dy \int_0^{1-y} dx \frac{y^{d/2 - 2 + \Delta}}{(y(1-y){q'}^2)^{1 + \Delta}}  \\
 & \qquad \times
 \sum_{n=0}^\infty \sum_{k=0}^n
 \binom{-1-\Delta}{n} \binom{n}{k} \frac{ (x(1-x)p^2 )^{n-k} (  - 2 x y p \cdot q' )^k }{(y(1-y){q'}^2)^n}  \, . \nonumber
\end{align}
We then rewrite as
\begin{align}
&\int d^d q' \frac{(p \cdot q')^k }{({q'}^2)^{2 + n +  \Delta} (p - q' )^2} = \frac{\Gamma(3 + n + \Delta)}{\Gamma (2 + n + \Delta)} \int d^d q' \int_0^1 dz \frac{(p \cdot q')^k (1-z)^{1 + n + \Delta}}{((q' - z p)^2 + z(1-z) p^2)^{3+n+\Delta}} \nonumber \\
&= \frac{\Gamma(3 + n + \Delta)}{\Gamma (2 + n + \Delta)} \int d^d w \int_0^1 dz 
 \sum_{l=0}^k \binom{k}{l} \frac{(p \cdot w)^l (z p^2)^{k-l} (1-z)^{1 + n + \Delta}}{(w^2 + z(1-z) p^2)^{3+n+\Delta}} \, .
\end{align}
For odd $l$, the integral over $w$ vanishes. 
For even $l$, we have
\begin{align}
\int d^d w \frac{(p \cdot w)^l}{(w^2 + z(1-z)p^2)^{3 + n + \Delta}} \propto
 \frac{\Gamma (3 + n + \Delta - d/2 - l/2)}{\Gamma (3 + n + \Delta)}
\frac{(p^2)^{l/2}}{(z(1-z)p^2)^{3 + n + \Delta - d/2 - l/2}} \, .
\end{align}
The integrals over $x,y,z$ are
\begin{align}
&\int_0^{1-y} dx x^n (1-x)^{n-k} 
 = \sum_{m=0}^{n-k} \binom{n-k}{m} (-1)^m \frac{(1-y)^{n+m+1}}{n+m+1} \, ,\nonumber \\
&\int_0^1 dy y^{d/2 - 3  - n + k} (1-y)^{ m - \Delta} =
\frac{\Gamma(d/2 - 2  - n + k ) \Gamma (m - \Delta + 1)}{\Gamma (m - n + k -  \Delta + d/2 - 1)}  \, , \\
&\int_0^1 dz z^{d/2 - 3-n-\Delta + k - l/2} (1-z)^{d/2 - 2 + l/2} =
\frac{\Gamma(d/2 - 2- n - \Delta + k - l/2) \Gamma (d/2 - 1 + l/2)}{\Gamma (d - 3 - n -  \Delta + k)}   \, . \nonumber 
\end{align} 
Therefore, after taking product of the three integrals, we do not have any term proportional to $1/\Delta$.

One of the integral of the type $K_1^{(2)}$ is $K _{k,l,m,n}$ defined in \eqref{Kklmn}.
We first evaluate the integral involving $q$ as
\begin{align}
\nonumber
&\int d^d q \frac{(\hat p - \hat q)^k  ( \hat q )^l   }{q^2  ((q - q')^2)^{\delta +  \Delta}}
= \frac{\Gamma(d/2 + \Delta)}{\Gamma(1)\Gamma(\delta  + \Delta)}\int d^d q  \int_0^1 dx \frac{(\hat p - \hat q)^k  ( \hat q )^l  x^{d/2 -2 + \Delta}}{((q - x q')^2 + x(1-x){q'}^2)^{d/2 + \Delta}} \\
& \quad =\pi^{d/2} \frac{\Gamma (\Delta)}{\Gamma (\delta + \Delta)} \int_0^1 dx \frac{(\hat p - x \hat q')^k  ( x \hat q' )^l  x^{d/2 -2  + \Delta}}{(x (1-x) {q'}^2)^{\Delta}} \, . 
\end{align}
Thus the $1/\Delta$-pole term becomes
\begin{align}
&\frac{1}{\Delta} \frac{ \pi^{d/2}}{\Gamma (\delta )} \sum_{a = 0}^k \binom{k}{a} (-1)^a (\hat p)^{k -a}  (\hat q ')^{a+l} 
 \int_0^1 dx  x^{a + l + d/2 - 2} \\
  & \qquad \qquad  \qquad \qquad \qquad 
  = \frac{1}{\Delta} \frac{ \pi^{d/2}}{\Gamma (\delta )}  \sum_{a = 0}^k \binom{k}{a} 
  \frac{(-1)^a (\hat p)^{k -a}  (\hat q ')^{a+l}}{a + l + \delta} \, . \nonumber
\end{align}
The integral over $q'$ is
\begin{align}
 \int d ^d q' \frac{(\hat q')^{a+l+m} (\hat p - \hat q')^n}{{q'}^2 ( p - q' )^2}
 & = \int d^d q ' \int_0^1 dy \frac{(\hat q')^{a+l+m} (\hat p - \hat q')^n}{((q' - y p)^2 + y(1-y) p^2)^2} \\
 & = \int_0^1 dy \pi^{d/2} \frac{\Gamma (2 - d/2)}{\Gamma(2)} \frac{1}{(y(1-y)p^2)^{2 - d/2}} ( y \hat p)^{a+l+m} (\hat p  - y \hat p)^n \nonumber \\
  &= \frac{\pi^{d/2} \Gamma(2 - d/2)\Gamma (a+l+m+\delta) \Gamma (n+\delta)}{\Gamma (a+l+m+n+2 \delta)} \frac{(\hat p)^{a+l+m+n}}{(p^2)^{2-d/2}}  \, . \nonumber
\end{align}
As we will show below, there is no $1/\Delta$-pole arising when exchanging the order of integrals over $q$ and $q'$.
Therefore we arrive at
\begin{align}
  \label{K12}
K _{k,l,m,n}
= \frac{ \pi ^d }{\Delta} \frac{\Gamma(2 - d/2)}{\Gamma(\delta)} \sum_{a = 0}^k \binom{k}{a} 
  \frac{(-1)^a }{a + l + \delta} \frac{\Gamma (a+l+m+\delta) \Gamma (n+\delta)}{\Gamma (a+l+m+n+2 \delta)}  + \mathcal{O} (\Delta^0)
\end{align}
in the momentum representation.

Let us show that there is no extra contribution by changing the order of integrals. In order to do so we compute
\begin{align}
 \int d^d q d^d q' \frac{1}{(p- q ')^2 {q}^2 (p-q )^2 ((q - q')^2)^{d/2 - 1 + \Delta}} \, , 
\end{align}
where $q,q'$ are replaced by $p-q' , p-q$.  The effects of derivatives  are neglected for simplicity, but it is not so difficult to includes them.
Then the integral over $q$ is given by \eqref{K11q}, and we can see there is no pole at $\Delta = 0$.
The $q'$-integral can be evaluated as before and we find no $1/\Delta$-pole term.

There are similar integrals with $p-q$ replaced by $q$ , $p-q'$ and $q'$.
We can show that these integrals are related to $K_{k,l,m,n}$  as 
\begin{align}
& \int d^d q d^d q' \frac{(\hat p - \hat q)^k  ( \hat q )^l   ( \hat q' )^m (\hat p - \hat q')^n }{(p-q)^2 {q'}^2 (p - q' )^2 ((q - q')^2)^{d/2 - 1 + \Delta}} = K_{l,k,n,m}  \frac{(\hat p)^{k+l+m+n}}{(p^2)^{2 - d/2}}\, , \nonumber \\
& \int d^d q d^d q' \frac{(\hat p - \hat q)^k  ( \hat q )^l   ( \hat q' )^m (\hat p - \hat q')^n }{(p-q)^2 q^2 {q'}^2 ((q - q')^2)^{d/2 - 1 + \Delta}} = K_{n,m,l,k}\frac{(\hat p)^{k+l+m+n}}{(p^2)^{2 - d/2}}\, , \label{exchange}\\
& \int d^d q d^d q' \frac{(\hat p - \hat q)^k  ( \hat q )^l   ( \hat q' )^m (\hat p - \hat q')^n }{(p-q)^2 q^2  (p - q' )^2 ((q - q')^2)^{d/2 - 1 + \Delta}} = K_{m,n,k,l}\frac{(\hat p)^{k+l+m+n}}{(p^2)^{2 - d/2}}  \nonumber
\end{align}
by exchanging $q$ and $q'$ or $q$ and $p-q$.

For the type $K_1^{(3)}$, we evaluate the integral $B_{k,l,m,n}$ defined in \eqref{Bklmn}.
We just need to repeat the analysis for \eqref{Bklsum} in the case of O$(N)$ scalars. The result is
\begin{align}
B_{k,l,m,n} 
& = \frac{2\pi^d}{\Delta} \frac{\Gamma(2 -d/2)}{\Gamma(d/2 - 2)} 
\sum_{a=0}^k \sum_{b=0}^l \binom{k}{a} \binom{l}{b} \frac{(-1)^a}{a+b+d/2-2}
\frac{\Gamma(k+b+d/2-1) \Gamma(l - b + 1)}{\Gamma (k + l + d/2)} \nonumber \\
& \times \frac{\Gamma(n-1+d/2) \Gamma(m+a+b-1+d/2)}{\Gamma(n+m+a+b-2 + d)} + \mathcal{O}(\Delta^0) \, .
\label{B12}
\end{align}


\providecommand{\href}[2]{#2}\begingroup\raggedright\endgroup

\end{document}